\PassOptionsToPackage{dvipsnames}{xcolor}
\documentclass[sigconf]{acmart}

\makeatletter
\def\@ACM@checkaffil{
    \if@ACM@instpresent\else
    \ClassWarningNoLine{\@classname}{No institution present for an affiliation}%
    \fi
    \if@ACM@citypresent\else
    \ClassWarningNoLine{\@classname}{No city present for an affiliation}%
    \fi
    \if@ACM@countrypresent\else
        \ClassWarningNoLine{\@classname}{No country present for an affiliation}%
    \fi
}
\makeatother

\copyrightyear{2023}
\acmYear{2023}
\setcopyright{acmlicensed}
\acmConference[ICS '23]{2023 International
Conference on Supercomputing}{June 21--23, 2023}{Orlando, FL, USA}
\acmBooktitle{2023 International Conference on Supercomputing (ICS '23), June
21--23, 2023, Orlando, FL, USA}
\acmPrice{15.00}
\acmDOI{10.1145/3577193.3593717}
\acmISBN{979-8-4007-0056-9/23/06}

\usepackage{tikz}
    \usetikzlibrary{positioning}
    \usetikzlibrary{arrows}
    \usetikzlibrary{backgrounds}
    \usetikzlibrary{arrows.meta}
\hypersetup{colorlinks,breaklinks,
            urlcolor=NavyBlue,
            linkcolor=NavyBlue}
\usepackage{amsmath}
\newcommand{\R}{\mathbb{R}}
\usepackage[ruled, linesnumbered, norelsize]{algorithm2e}
\usepackage{array}
\usepackage{booktabs, multirow}
\usepackage{enumitem}
\usepackage{graphicx}
\usepackage{xcolor}
\definecolor{crimson}{rgb}{0.86, 0.08, 0.24}

\usepackage{listings}
\lstdefinestyle{C-plus-plus}{
	belowcaptionskip=1\baselineskip,
	breaklines=true,
	xleftmargin=\parindent,
	language=C++,  
	showstringspaces=false,
	basicstyle=\footnotesize\ttfamily,
	keywordstyle=\bfseries\color{PineGreen},
	commentstyle=\color{white!40!black},
	identifierstyle=\color{NavyBlue},
	stringstyle=\color{Orange},
	escapeinside={(*@}{@*)},
	morekeywords={pragma, omp, parallel, shared},
	numbers=left,
	morekeywords={[2]{row, Zero}},
	keywordstyle={[2]{\color{Purple}}},
}  

\usepackage{caption}
\usepackage{subcaption}
\newcommand{\thiswork}{HEAT}

\begin{document}

\title[\thiswork]{\thiswork: A Highly Efficient and Affordable Training System for Collaborative Filtering Based Recommendation on CPUs}

\newcommand{\iu}{Indiana University}
\newcommand{\anl}{Argonne National Labotary}
\newcommand{\microsoft}{Microsoft}

\newcommand{\AFFIL}[4]{%
    \affiliation{
        \institution{\small #1}
        \city{#2}\state{#3}\country{#4}
    }
    }

\newcommand{\IU}{\AFFIL{\iu}{Bloomington}{IN}{USA}}
\newcommand{\ANL}{\AFFIL{\anl}{Lemont}{IL}{USA}}
\newcommand{\MS}{\AFFIL{\microsoft}{Redmond}{WA}{USA}}

\settopmatter{authorsperrow=5}

\author{Chengming Zhang}{\IU}
\email{czh5@iu.edu}

\author{Shaden Smith}{\MS}
\email{shaden.smith@microsoft.com}

\author{Baixi Sun}{\IU}
\email{sunbaix@iu.edu}

\author{Jiannan Tian}{\IU}
\email{jti1@iu.edu}

\author{Jonathan Soifer}{\MS}
\email{jonso@microsoft.com}

\author{Xiaodong Yu}{\ANL}
\email{xyu@anl.gov}

\author{Shuaiwen Leon Song}{\MS}
\email{leonsong@microsoft.com}

\author{Yuxiong He}{\MS}
\email{yuxhe@microsoft.com}

\author{Dingwen Tao}{\IU}
\authornote{Corresponding author: Dingwen Tao, Department of Intelligent Systems Engineering, Luddy School of Informatics, Computing, and Engineering, Indiana University.
}
\email{ditao@iu.edu}

\renewcommand{\shortauthors}{Zhang et al.}

\begin{abstract}
Collaborative filtering (CF) has been proven to be one of the most effective techniques for recommendation. Among all CF approaches, SimpleX is the state-of-the-art method that adopts a novel loss function and a proper number of negative samples. However, there is no work that optimizes SimpleX on multi-core CPUs, leading to limited performance. To this end, we perform an in-depth profiling and analysis of existing SimpleX implementations and identify their performance bottlenecks including (1) irregular memory accesses, (2) unnecessary memory copies, and (3) redundant computations. To address these issues, we propose an efficient CF training system (called \thiswork) that fully enables the multi-level caching and multi-threading capabilities of modern CPUs. Specifically, the optimization of \thiswork{} is threefold: (1) It tiles the embedding matrix to increase data locality and reduce cache misses (thus reduces read latency); (2) It optimizes stochastic gradient descent (SGD) with sampling by parallelizing vector products instead of matrix-matrix multiplications, in particular the similarity computation therein, to avoid memory copies for matrix data preparation; and (3) It aggressively reuses intermediate results from the forward phase in the backward phase to alleviate redundant computation. Evaluation on five widely used datasets with both x86- and ARM-architecture processors shows that \thiswork{} achieves up to 45.2$\times$ speedup over existing CPU solution and 4.5$\times$ speedup and 7.9$\times$ cost reduction in Cloud over existing GPU solution with NVIDIA V100 GPU.
\end{abstract}


\maketitle
\pagenumbering{gobble}

\section{Introduction}
\label{sec:introduction}
With the development and popularization of the Internet and smart devices, these platforms have become ideal tools for collecting various data/information that can be used to identify user preferences \cite{ko2022survey}. However, the exponential growth in the amount of digital information and the explosion in the number of Internet users can lead to the problem of information overload that prevents timely access to things of interest on the Internet \cite{isinkaye2015recommendation}. As a result, recommender systems (a.k.a recommendation systems) that use a user's choices, interests, or observed behavior to filter out key information from a massive amount of dynamically collected information are more in demand than ever \cite{pan2010research}.

Collaborative filtering (CF) has been proven to be one of the most effective techniques for building recommender systems due to its ability to recommend completely dissimilar content. Learning effective latent factors directly from the user-item rating matrix through matrix factorization (MF) is the most effective method among CF-based approaches \cite{koren2009matrix} (will be discussed in \S\ref{subsec:MF}). Despite the effectiveness of MF-based CF, training MF-based CF is challenging due to two reasons: 
(1) \textbf{Performance:} Irregular memory access causes a significant performance degradation when using sparse real-world user-item rating matrices \cite{dong2017hybrid}. 
(2) \textbf{Cost:} Weekly or monthly updates on large training datasets (millions of users and items) causes a drastic cost increase. 
Recommendation tasks in the industry are time-sensitive and profit-oriented, which means that training time and cost are critical for entrepreneurs.

Although today's various accelerators such as graphics processing units (GPUs) are widely used to train machine learning models today, in this work we focus on using multi-core CPUs for training recommendation models due to the following three main reasons.

\textbf{CF training with GPUs is expensive}. Nowadays, users typically accelerate machine learning applications \cite{steinkraus2005using} using GPUs that are capable of performing floating-point operations in a massively parallel fashion. However, CF training on large datasets using GPUs is expensive. For example, training an MF-based CF model for 3,000 epochs on a dataset with 200 million users and 200 million items takes about 297.7 hours using 100 16 GB V100 GPUs. Assuming this training service is deployed on an AWS p3.2xlarge instance \cite{p3.2xlarge} at an hourly cost of around 3 dollars, the total cost for one-time training is approximately 91,081 dollars. Furthermore, the model needs to be retrained every month or week due to the dataset update.

\textbf{GPU memory is limited}. Besides the user-item rating matrix as input, training MF-based CF also requires holding two large embedding matrices in memory, i.e., a user embedding matrix $S \in \R ^{|U| \times K}$ and an item embedding matrix $T \in \R ^{|I| \times K}$. Here $|U|$ is the number of users, $|I|$ is the number of items, and $K$ is usually between 64 and 128. 
For instance, the Amazon Product Reviews dataset \cite{haque2018sentiment, he2016ups} contains 21 million users and 9 million items. However, currently the most powerful GPUs have only 80 GB of memory \cite{choquette2021nvidia}, which can only accommodate embedding matrices of up to 4 million users/items when $K$ is 128 and the data type is float32. For larger datasets, we need to split embedding matrices and rating matrix across different GPUs, which introduces high global communication overhead. In comparison, the popular CPU nodes in HPC systems typically have 512 GB or even 1-2 TB of memory.


\textbf{MF-based CF is not suitable for GPUs}. An MF-based CF features low computation intensity and highly irregular memory accesses. For MF-based CF training, we need to maximize the similarity of a positive user-item pair while minimizing the similarity of a negative user-item pair. This procedure requires fetching (1) a $K$-dimensional user embedding (vector) from the user embedding matrix, (2) one $K$-dimensional positive item embedding from the item embedding matrix, and (3) random $n$ $K$-dimensional negative item embeddings from the item embedding matrix. Based on these sampled vectors, we need to compute the similarity of the user-item pairs using vector-dot product operations. These computation patterns (i.e., embeddings are first accessed in an irregular fashion and then used in a regular way with spatial locality for low computation-intensive vector products) make MF-based CF training more suitable for CPUs than GPUs.

Prior works have focused on optimizing the performance of MF-based CF training. 
For example, MSGD \cite{li2017msgd} improves training performance on GPUs by removing dependencies on user and item pairs. However, MSGD does not support sampling multiple negative terms, which leads to inferior training results (i.e., low accuracy). 
Recently, Mao \textit{et al.} proposed SimpleX \cite{mao2021simplex}, a state-of-the-art CF method, that has a novel loss function and a large negative sampling rate, greatly outperforming other existing methods. However, SimpleX only uses PyTorch to implement its approach without considering the computational efficiency. 
Specifically, (1) training on sparse user-item rating matrices and random sampling for multiple negative items lead to irregular memory accesses to embedding matrices.
(2) The similarity computation before the loss computation is usually based on parallel matrix-matrix multiplication, which introduces expensive memory copies to concatenate sampled vectors into matrices.
(3) Automatic differentiation engines in machine learning frameworks (such as \texttt{autograd} in PyTorch) ignore potential data reuse in the backward phase (see \S\ref{subsec:reuse}).

To this end, we propose a \underline{H}ighly \underline{E}fficient and \underline{A}ffordable \underline{T}raini\-ng system (called \thiswork\footnote{The code is available at \url{https://github.com/hipdac-lab/HEAT}.}) for collaborative filter based recommendation on multi-core CPUs based on the SimpleX approach.
First, we propose to take advantage of modern CPUs' memory hierarchies to reduce embedding read latencies. Specifically, we propose an effective tiling method that partitions item embedding matrices to fit into multi-level caches according to their sizes.
Second, we adopt a multi-threaded training method \cite{recht2011hogwild}, where each thread independently and parallelly reads its corresponding user and item embeddings, calculates their gradients, and updates them rather than all embeddings, and fuse forward and backward phases to reduce the size of the per-thread memory footprint.
Third, we identify reuse opportunities for intermediate results during the backward pass of training. This reuse is missed in automatic differentiation systems that work at a more fine-grained operator level.
To the best of knowledge, \textit{this is the first work that enables high-performance and low-cost CF training for recommendation based on the SimpleX approach on multi-core CPUs}. 

The main contributions of this paper are summarized as follows:

\begin{itemize}[noitemsep, topsep=2pt, leftmargin=1.3em]
  \item We deeply analyze the performance of two state-of-the-art MF-based CF solutions and identify their performance bottlenecks.
  \item We propose to tile the item embedding matrix according to multi-level cache sizes to reduce read latency. Furthermore, we propose a light-weight algorithm to find the optimal tiling size and cache eviction policy (e.g., refresh interval).
  \item We develop a parallel method for similarity computation based on vector products rather than matrix-matrix multiplication to avoid matrix data preparation (i.e., memory copies).
  \item We propose to save the result of the partial derivative of embeddings in the forward computation and reuse them in the backward computation to avoid redundant calculations.
  \item We propose two implementation optimizations to improve the performance of weight updates by alleviating read/write conflicts in shared memory and reducing the amount of updates.
  \item Evaluation on three real-world datasets with AMD 7742 CPUs and Fujitsu A64FX CPUs shows that \thiswork{} achieves up to 45.2$\times$ and 4.5$\times$ speedups over state-of-the-art CPU and GPU solutions, respectively. We also derive some takeaways for CF training on different CPU architectures.
\end{itemize}

The remaining of the paper is organized as follows.
In \S\ref{sec:background}, we present the background about recommender systems and matrix factorization based collaborative filtering. In \S\ref{sec:motivation}, we present our profiling and analysis of existing solutions. In \S\ref{sec:design}, we describe the design of our \thiswork. In \S\ref{sec:evaluation}, we evaluate \thiswork{} on different datasets and compare it with other works. In \S\ref{sec:related_work}, we discuss related work. In \S\ref{sec:conclusion}, we conclude our work and discuss future work.
\section{Background}
\label{sec:background}
In this section, we present the background information about recommender systems and collaborative filtering techniques for recommendation.

\subsection{Basics of Recommender Systems}
\textbf{Input Data.} There are two main types of collected feedback (i.e., rating matrix): (1) explicit feedback \cite{koren2009collaborative} that is directly provided by users, such as likes and ratings, and (2) implicit feedback \cite{bayer2017generic} that is obtained from users' interactions, such as click data, purchases, and implicit visit information. Recent research on recommender systems has shifted from explicit feedback to implicit feedback \cite{ding2018improving} because the majority of a user's preference-related signal is implicit. Thus, we focus on implicit feedback in this work.

\textbf{Filtering Techniques.} Recommender systems mainly include content-based \cite{cantador2008multi} filtering and collaborative filtering (CF) \cite{im2007does} techniques. Content-based filtering is based on the items' information and recommends items that have attributes similar to those that users like. However, the technique is notorious for its inability to recommend dissimilar items \cite{salter2006cinemascreen}. To address this issue, collaborative filtering makes recommendations by learning preferences or taste information from many other users’ interactions \cite{im2007does} and is able to provide diverse recommendations. CF techniques can be further classified into user–user CF \cite{resnick1994grouplens}, item-item CF \cite{sarwar2001item}, dimensionality reduction \cite{billsus1998learning}, and probabilistic methods \cite{kitts2000cross}.
Specifically, user–user and item–item CF techniques directly use feedback to calculate similarities between users or items. But vectors in feedback are highly sparse and have extremely large dimensions. For example, an item is a $|U|$-dimensional vector and a user is a $|I|$-dimensional vector, where $U$ is the set of all users and $I$ is the set of all items. This causes high overhead of computing resources and memory space.
To address this issue, dimensionality reduction techniques such as matrix factorization (MF) reduce the dimension of the rating space to a constant number $K$ \cite{schafer2007collaborative} thereby reducing computational complexity and memory requirements (will be detailed in the next section).
Other techniques such as probabilistic methods seek to create probabilistic models of users' behaviors and employ those models to predict users' future behaviors.

\textbf{Software Frameworks}. There are two popular training frameworks to implement recommender systems: 
(1) PyTorch provides a lookup table (\texttt{torch.nn.Embedding}) to store embeddings of a fixed dictionary and size. Users can build a complete model using necessary modules (e.g., loss function, optimizer) provided by PyTorch. Besides, users can utilize the \texttt{autograd} module to compute gradients and then update the embeddings. However, on the one hand, using \texttt{torch.nn.Embedding} with dense gradient would directly update all embeddings, which leads to unnecessary operations since only part of the embeddings are involved in one training iteration; on the other hand, using \texttt{torch.nn.Embedding} with sparse gradient causes worse performance (detailed in Section \S\ref{subsec:update}).
(2) TorchRec \cite{ivchenko2022torchrec} is a production-quality recommender systems package in the open-source PyTorch ecosystem. It provides model and data parallelism and represents sparse inputs by jagged tensors. Moreover, TorchRec supports computations on sparse data through FBGEMM \cite{khudia2021fbgemm} and overlaps communication and computation through train\_pipeline. However, similar to PyTorch, TorchRec also suffers from the above dense/sparse embedding update issue. 

\begin{figure}[t]
    \centering
    \includegraphics[width=\linewidth]{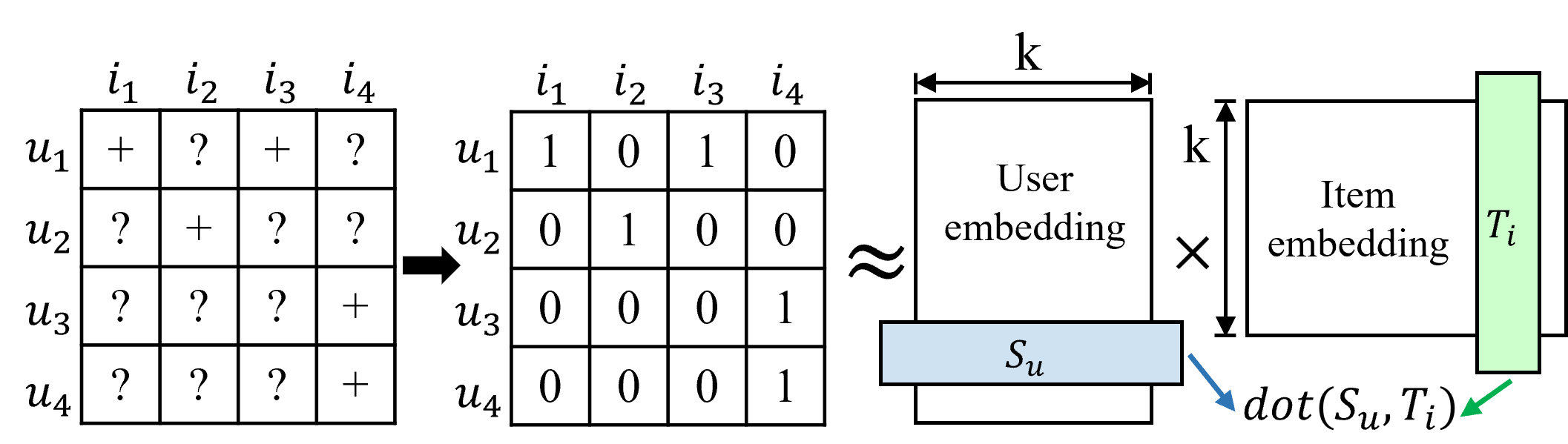}
    \caption{Basic concept of MF-based CF.}
    \label{fig:mf_basic}
\end{figure}

\subsection{MF-based Collaborative Filtering}
\label{subsec:MF}
The purpose of MF-based CF training is to maximize the similarity of embeddings of a positive user-item pair while minimizing the similarity of embeddings of a negative user-item pair. We can use dot product similarity or cosine similarity as expressed in Equation \ref{equ-2}.
Assume $U$ is the set of all users and $I$ is the set of all items. The implicit feedback can be expressed as $X \subseteq U \times I$ as depicted in Figure~\ref{fig:mf_basic}. Particularly, ``+'' indicates a user's preference for an item. Such corresponding items are called \textbf{positive items}. ``?'' represents either negative (not interested) or missing (not interacted) values. The items corresponding to the negative values are called \textbf{negative items}. MF-based techniques train two low-dimensional matrices, i.e., a user embedding matrix $S \in \R ^{|U| \times K}$ and an item embedding matrix $T \in \R ^{|I| \times K}$, to approximate $X$ as expressed in Equation (\ref{equ-1}). Then, the main task is to predict missing ratings in $X$ using the corresponding embeddings.

{\small
\begin{align}
    X \approx & \ \hat{X} = ST^t
    \label{equ-1} \\
    \hat{x}_{u, i} = & \left\{
    \begin{aligned}
         & S_u \cdot T_i = \sum_{k=0}^{K} S_{u, k} T_{i, k} \quad\text{(dot)} \\
         & \frac{S_u \cdot T_i}{\|S_u\|_2 \|T_i\|_2} = \frac{\sum_{k=0}^{K} S_{u, k} T_{i, k}}{\sqrt{\sum_{k=0}^{K} S^2_{u, k}} \sqrt{\sum_{k=0}^{K} T^2_{i, k}}} \quad\text{(cosine)} 
    \end{aligned}
    \right.
    \label{equ-2} \\
    \mathcal{L}(u, i) & = (1 - \hat{x}_{u, i}) + \frac{\mu}{|\mathcal{N}|} \sum_{j \in \mathcal{N}} \text{max}(0, \: \hat{x}_{u, j} - \theta)
    \label{equ-6} 
\end{align}
}

Prior MF-based CF works can be generally classified into two directions. The first direction only targets recall (i.e., accuracy) and adopts simple similarity functions (e.g., dot product) and point-wise loss functions (e.g., mean square error, binary cross entropy) for user-item pairs without using negative items. For example, representative works such as CuMF\_ALS \cite{tan2016faster}, CuMF\_SGD \cite{xie2016cumf_sgd}, and MSGD \cite{li2017msgd} focus more on the computational efficiency of matrix factorization than on recall.
The second direction brings higher accuracy and creates a user-specific item ranking by using the concept of positive/negative items, novel loss functions, and more sophisticated similarity functions (e.g., cosine similarity) with sampling methods. For example, \cite{rendle2012bpr} proposed Bayesian personalized ranking (BPR) loss function, while \cite{hadsell2006dimensionality, wang2021understanding} proposed a contrastive loss (i.e., a Euclidean distance-based loss). More recently, SimpleX \cite{mao2021simplex} proposes a cosine contrastive loss (CCL) and utilizes \textit{multiple} negative samples, which outperforms other approaches regarding accuracy. Equation (\ref{equ-6}) is the CCL, where $(u, i)$ is a positive user-item pair, $\mathcal{N}$ is the number of randomly sampled negative samples, $\mu$ is a hyperparameter, and $\theta$ is the threshold to filter negative samples.




\section{Performance Profiling \& Analysis}
\label{sec:motivation}
In this section, we characterize the performance of SimpleX on both CPU and GPU.
Note that we focus on the PyTorch implementation rather than the TorchRec implementation since TorchRec optimizes sparse computation and communication/computation overlap, which does not address the performance bottleneck of MF-based CF training methods. Thus, for simplicity, we only show the performance breakdown of the PyTorch implementation to motivate the design of our \thiswork.

We set the embedding dimension to 128 and the number of negatives to 64, and the batch size to 1024. We use PyTorch 1.10.0 and CUDA 10.2. We use the PyTorch profiler \cite{profiler} to perform the breakdown. We use three real-world large datasets (containing more than millions of users and items) for profiling, i.e., Goodreads Book Reviews (Goodreads) \cite{wan2018item}, Google Local Reviews (2018) (Google) \cite{pasricha2018translation}, and Amazon Product Reviews (Amazon) \cite{he2016ups}.

\begin{figure}[t]
    \centering
    \resizebox{.8\linewidth}{!}{
    {\small
        \begin{tikzpicture}

\tikzset{Shape/.style={draw}};
\tikzset{Stripe/.style={draw, inner sep=0pt, minimum height=.07in, minimum width=.9in}};
\tikzset{StripeVert/.style={draw, inner sep=0pt, minimum width=.07in, minimum height=.9in}};

\begin{scope}

\begin{scope}[xshift=-0.8in]

\node[fill=white, align=center, Shape, minimum height=1in, minimum width=.8in] (user-matrix) at (0,0) {User\\embedding};

\draw[|{latex}-{latex}|] ([yshift=1ex]user-matrix.north west) -- node[above]{$K$} ([yshift=1ex]user-matrix.north east);
\draw[|{latex}-{latex}|]  
([xshift=-1.5ex]user-matrix.south west)
-- node[sloped, above]{\#user} 
([xshift=-1.5ex]user-matrix.north west);

\foreach \y/\pn in {-0.9/+, 0.8/+} {

    \node[fill=none, very thick, draw=white, align=center, Stripe] at (0,\y) {};
    
    \node[fill=NavyBlue, fill opacity=.2, Stripe] at (0,\y) {};
}

\end{scope}

\begin{scope}[xshift=.8in]

\node[fill=white, align=center, Shape, minimum height=1in, minimum width=.8in] (item-matrix) at (0,0) {Item\\embedding};

\draw[|{latex}-{latex}|] ([yshift=1ex]item-matrix.north west) -- node[above]{$K$} ([yshift=1ex]item-matrix.north east);
\draw[|{latex}-{latex}|]  
([xshift=-1.5ex]item-matrix.south west)
-- node[sloped, above]{\#item} 
([xshift=-1.5ex]item-matrix.north west);

\foreach \y/\color/\pn in {-1.1/green/-, -0.8/green/-, -0.5/green/+, 0.5/green/-, 0.8/green/-, 1.1/green/+} {

    \node[fill=none, very thick, draw=white, align=center, Stripe] at (0,\y) {};
    
    \node[fill=\color, fill opacity=.2, Stripe] at (0,\y) {};
    \node[font=\tiny] at (0,\y) {($\pn$)};
}

\end{scope}

\end{scope}


\begin{scope}[yshift=-1.5in]
    
\begin{scope}[xshift=-.7in]
    
\node[StripeVert, fill=NavyBlue, fill opacity=.2] (b-1) {};
\node[StripeVert, fill=NavyBlue, fill opacity=.2, right=1pt of b-1] (b-2) {};

\draw[|{latex}-{latex}|] ([yshift=1ex]b-1.north west) -- node[above] (middle-anchor) {\#batch} ([yshift=1ex]b-2.north east);
\draw[|{latex}-{latex}|]  
([xshift=-1.5ex]b-1.south west)
-- node[sloped, above]{$K$} 
([xshift=-1.5ex]b-1.north west);

\end{scope}

\begin{scope}[xshift=.6in]

\node[Stripe, fill=green, fill opacity=.2, minimum height=.5in] (sample) {};
\node[Stripe, fill=none, draw=white, very thick, minimum height=.5in, xshift=.08in, yshift=-.08in] (sample-2) {};
\node[Stripe, fill=green, fill opacity=.2, minimum height=.5in, xshift=.08in, yshift=-.08in] {};

\draw[|{latex}-{latex}|] ([yshift=1ex]sample.north west) -- node[above] (middle-anchor-R) {$K$} ([yshift=1ex]sample.north east);
\draw[|{latex}-{latex}|]  
([xshift=-1.5ex]sample.south west)
-- node[sloped, above]{\#pos-neg-sample} 
([xshift=-1.5ex]sample.north west);

\draw[|-|] (sample.north east) -- node[sloped, above] (middle-anchor-R) {\#batch} (sample-2.north east);
    
\end{scope}

\end{scope}

\draw[-latex] (user-matrix.south) -- (middle-anchor);
\draw[-latex] (item-matrix.south) -- (middle-anchor-R);



\node[rounded corners=4pt, minimum height=1.4in, minimum width=2.7in, draw] at (0, -1.4in) (bottom) {};

\node[fill=white] at (bottom.north){Read \& Reshape};

\node[rounded corners=4pt, draw, below=4ex of bottom, align=center] (output) {Normalization and \\matrix-matrix multiplication};

\draw[-latex] (bottom) -- (output);

\end{tikzpicture}
    }
    }
    \caption{Overview of SimpleX. ``+''/``-'' denote positive/negative embedding.}
    \label{fig:SimpleX_framework}
    \vspace{-2mm}
\end{figure}
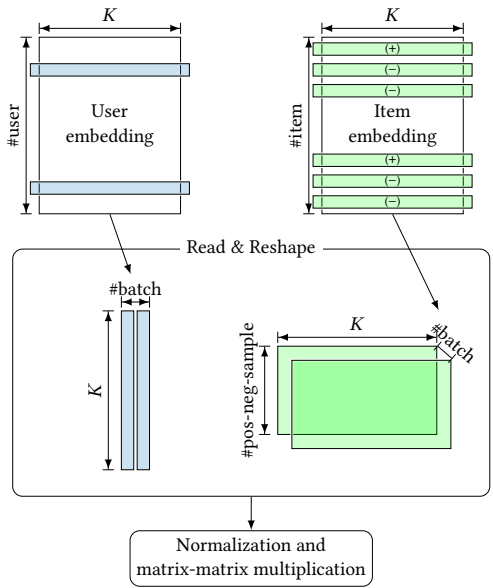

\begin{figure*}[ht]
    \centering
    \includegraphics[width=\linewidth]{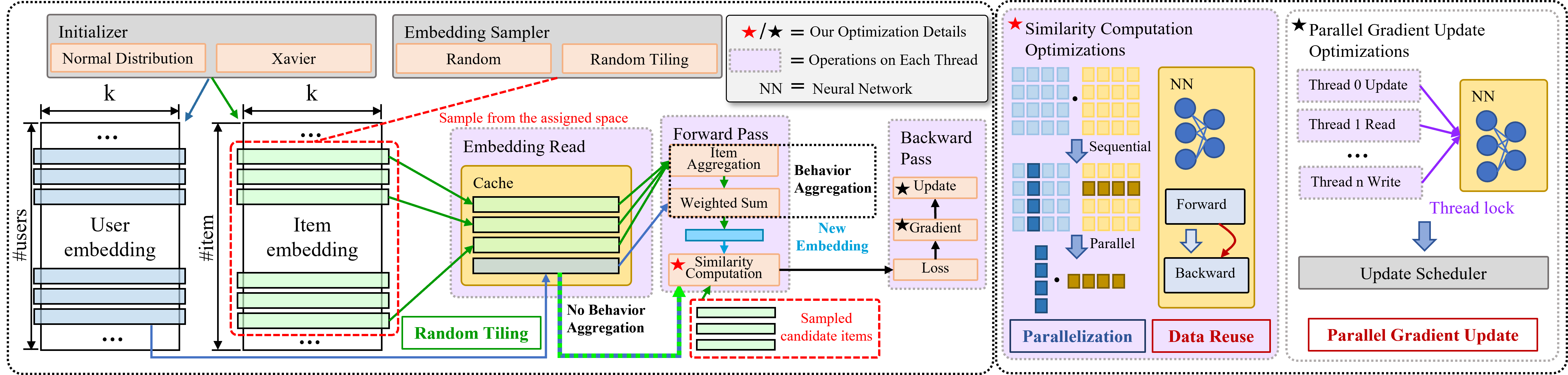}
    \caption{Overview of our proposed \thiswork's workflow/dataflow. HEAT has four main optimizations: ``random tiling'' (\S\ref{subsec:tiling}), ``parallelization'' (\S\ref{subsec:similarity}), ``data reuse'' (\S\ref{subsec:reuse}) and ``parallel gradient update'' (\S\ref{subsec:aggregator}).}
    \vspace{-2mm}
    \label{fig:overview}
\end{figure*}

\subsection{Embedding Update in SimpleX}
\label{subsec:update}
The core component in the PyTorch implementation of SimpleX \cite{mao2021simplex} is \texttt{torch.nn.Embedding}, a simple lookup table storing embeddings. SimpleX fetches a batch of embeddings from torch.nn.Embedding to perform one training iteration. Logically, we only need to generate the gradients of the involved embeddings and update those embeddings. Thus, we can leverage \texttt{torch.nn.Embedding}'s capability which allows users to enable sparse gradient computation and embedding update (by setting the parameter \texttt{sparse} to True).

\begin{table}[t]
    \caption{Profiling of embedding update in SimpleX. ``ET'', ``FP'', ``BP'' are short for epoch time, forward percentage, backward percentage, respectively.}
    \centering
    \resizebox{0.75\linewidth}{!}{
        \begin{tabular}{rrrrr}
\toprule
  \textbf{\sffamily Dataset}
& \textbf{\sffamily Method}
& \textbf{\sffamily ET}
& \textbf{\sffamily FP}
& \textbf{\sffamily BP} \\
\midrule
\multirow{2}{*}{AmazonBooks}&dense  &257.4  &19.9\%   &67.0\%\\
                            &sparse &946.6  &6.2\%    &92.8\%\\
\multirow{2}{*}{Yelp18}     &dense  &129    &21.2\%   &65.1\%\\
                            &sparse &386.3  &9.1\%    &89.3\%\\
\multirow{2}{*}{Gowalla}    &dense  &94.9   &20.7\%   &66.8\%\\
                            &sparse &251.8  &9.2\%    &89.2\%\\
\bottomrule
\end{tabular}
    }
    \label{tab:emb_update}
\end{table}

\begin{table}[b]
    \caption{Breakdown of the forward phase of SimpleX. The ratio of each component's time to the forward time. Amazon is short for AmazonBooks.}
    \centering
    \resizebox{0.9\linewidth}{!}{
        \begin{tabular}{rrrrrrrr}
\toprule
  \textbf{\sffamily Dataset}
& \textbf{\sffamily u\_emb}  
& \textbf{\sffamily i\_emb} 
& \textbf{\sffamily u\_norm} 
& \textbf{\sffamily i\_norm} 
& \textbf{\sffamily mem\_cp} 
& \textbf{\sffamily bmm} 
& \textbf{\sffamily loss}\\
\midrule
AmazonBooks  &9.6\%  &39.8\%  &5.9\%  &22.3\%  &5.0\%  &7.1\%  &9.7\%\\
Yelp18       &9.1\%  &35.3\%  &5.1\%  &28.3\%  &4.8\%  &7.2\%  &9.6\%\\
Gowalla      &8.3\%  &33.2\%  &5.6\%  &31.1\%  &4.8\%  &7.3\%  &9.1\%\\
\bottomrule
\end{tabular}

    }
    \label{tab:efficiency}
\end{table}

Table~\ref{tab:emb_update} shows the profiling results of the embedding update in SimpleX with dense or sparse gradients.
In the case of dense gradient, the backward phase takes more than 60\% of the epoch time. We observe that \texttt{torch.nn.Embedding} updates all embeddings in every iteration.
In the case of sparse gradient, although we theoretically reduce the computation complexity, the actual epoch time of training with sparse gradient is almost 3$\times$ higher than that of dense gradient, where the backward phase takes more than 90\% of the epoch time. 
This motivates us to design a training method that supports updating embedding sparsely and efficiently in parallel.

\subsection{Computation Efficiency of SimpleX}
\label{subsec:efficiency}
SimpleX utilizes \texttt{torch.bmm}, a batch matrix-matrix product for similarity computation. Before that, it needs to concatenate and then reshape embeddings. Specifically, as shown in Figure~\ref{fig:SimpleX_framework}, SimpleX reads a batch of user embeddings and item embeddings, and reshapes them to let batch dimension be the first dimension. After reshaping, SimpleX performs normalization and matrix-matrix multiplication. \texttt{torch.bmm} can fully enable the underlying high-performance BLAS library on multi-core CPUs.

Table~\ref{tab:efficiency} shows the breakdown of the forward phase of SimpleX. The forward phase includes reading user embeddings (u\_emb), reading item embeddings (i\_emb), normalization of user embeddings (u\_norm), normalization of item embeddings (i\_norm), concatenation and reshaping of embeddings (mem\_cp), a batch matrix-matrix product (bmm), and a loss function (loss). 
We observe that the time of mem\_cp and the time of bmm are comparable. This inspires us to avoid explicit concatenation and reshaping.
To normalize the embedding tensor $E$, it needs to sum the square of $E$ along the dimension of the embedding dimension, and then calculate the square root of the summation, and then reverse values of the square root, i.e., the norm $R = \left( \sqrt{E^2.sum(dim=1)} \right)^{-1}$. We observe that this normalization takes more than 20\% of the forward time because of two main reasons: (1) The underlying library does not have good support for the above operations. (2) Reading the entire matrix during the computation and the writing of the generated intermediate tensor $R$ back to the memory cause additional memory access time.
In addition, the time of reading item embeddings takes around 30\% of the forward time, which is caused by irregular memory accesses. 

\subsection{Memory Usage of SimpleX}
The sizes of user and item embedding matrices in MF-based CF are linearly scaled to the size of training dataset (i.e., item-user rating matrix). Table~\ref{tab:memory} shows the memory usage of SimpleX on both CPU and GPU. The total memory capacity of GPU and CPU is 32 GB and 256 GB, respectively. We observe that SimpleX almost runs out of the GPU memory when the numbers of users and items are over 3 millions. 
This is because it needs to save not only the embedding matrices, but also the gradient matrices (scaled with user/item sizes) and optimizer states (scaled with batch size).
The out of memory happens when training on the Amazon dataset due to the limited GPU memory. This observation further strengthens our motivation to use multi-core CPUs with larger memory as our target platform. 

\begin{table}[t]
    \caption{Memory usage of SimpleX. OoM is short for out of memory.}
    \centering
    \resizebox{0.6\linewidth}{!}{
        \begin{tabular}{rrrrr}
\toprule
  \textbf{\sffamily Dataset}
& \textbf{\sffamily users}  
& \textbf{\sffamily items} 
& \textbf{\sffamily CPU} 
& \textbf{\sffamily GPU}\\
\midrule
Goodreads &0.81M  &1.56M  &4.2\%   &30.1\%\\
Google    &4.57M  &3.12M  &11.3\%  &80.2\% \\
Amazon    &20.98M & 9.35M &38.4\%  &OoM\\
\bottomrule
\end{tabular}
    }
    \vspace{2mm}
    \label{tab:memory}
\end{table}

\section{Design Methodology}
\label{sec:design}
In this section, we propose our multi-threading MF-based CF training system with optimizations to improve the training performance.

\subsection{Overview of \thiswork}
\label{subsec:overview}
Figure~\ref{fig:overview} shows the key components of \thiswork{}: (1) It initializes user/item embedding matrices with values either drawn from the normal distribution $\mathcal{N}(mean, std^2)$ or initialized by Xavier \cite{glorot2010understanding}. (2) It chooses either the original random sampler or our proposed random tiling sampler that increases the cache hit ratio (see in \S\ref{subsec:tiling}) to sample one user, one positive item, and $n$ negative items. Then, it reads the user's corresponding embeddings and these positive/negative items' embeddings. 
(3) The behavior aggregation layer with our proposed optimization of gradient update (see \S\ref{subsec:aggregator}) generates a new user embedding via aggregating embeddings of historically interacted items of the user when enabling behavior aggregation. 
(4) It calculates similarities in parallel (see \S\ref{subsec:similarity}) of user-item pairs and calculate the loss. 
(5) It calculates gradients through an optimized gradient computation kernel (see \S\ref{subsec:reuse}). (6) It updates and writes back corresponding embeddings. 

\vspace{-1mm}
\subsection{Random Tiling}
\label{subsec:tiling}
\textbf{Cache size oriented tiling.} The original method randomly samples $n$ negative items following a uniform distribution and reads their embeddings. As shown in Table~\ref{tab:naive}, the time of reading item embeddings exceeds 60\% of the total forward time. This is due to two reasons: (1) randomly sampled negative items lead to irregular memory accesses, which causes poor data locality, low cache hit rate, and high latency. 
(2) Each embedding consists of $K (K \geq 64)$ floating-point numbers, which will further exacerbate this problem when $K$ is relatively large. Meanwhile, reading user embeddings takes less than 5\% of the total forward time because we only sample one user in each iteration.

\begin{table}[t]
    \caption{Breakdown of the forward phase of HEAT with random sampling.}
    \centering
    \resizebox{0.65\linewidth}{!}{
        \begin{tabular}{rrrrrrrr}
\toprule
  \textbf{\sffamily Dataset}
& \textbf{\sffamily u\_emb}  
& \textbf{\sffamily i\_emb} 
& \textbf{\sffamily compute} 
& \textbf{\sffamily loss}\\
\midrule
Amazon  &5.1\% &63.2\% &25.9\% &4.3\% \\
Yelp18  &5.3\% &62.4\% &26.6\% &4.6\% \\
Gowalla &5.5\% &61.4\% &26.4\% &4.5\% \\
\bottomrule
\end{tabular}
    }
    \label{tab:naive}
\end{table}

To utilize modern CPU's memory hierarchy, especially multi-level caches, we propose to tile the item embedding matrix according to the cache size and make sure a tile of items and their embeddings can be fitted into the cache. Then, we randomly sample negative items directly from the cached tile of items, which increases the cache hit rate and thus reduces the latency of reading embeddings.
Assume $N_1$ and $N_2$ are tiling size and refresh interval, respectively. The sampling space of the original strategy is whole items. However, the sampling space is shrunk to the tiling size $N_1$ after applying the tiling strategy. To reduce the impact of the tiling strategy on training results as much as possible, we hope to have as large a sampling space as possible while ensuring acceleration. Thus, we will refresh the cached tile every $N_2$ iterations to enlarge the sampling space. The sampling space becomes $\frac{M}{N_2} \times N_1$, where $M$ is the number of total iterations.

\begin{figure}[b]
    \centering
    \includegraphics[width=\linewidth]{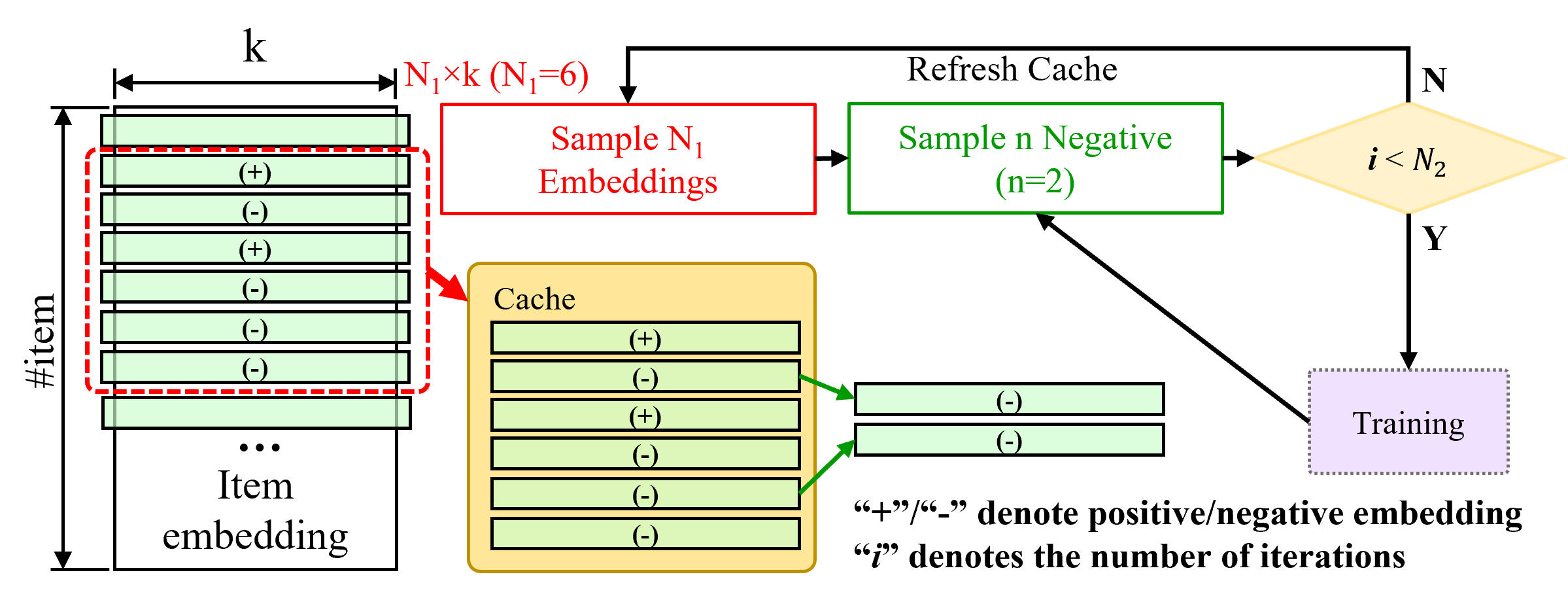}
    \caption{Random tiling strategy in each thread.}
    \label{fig:random_tile}
\end{figure}

Figure~\ref{fig:random_tile} illustrates the proposed random tiling strategy in each thread. Specifically, each thread preallocates a suitable cache space to buffer randomly sampled $N_1$ embeddings. In each iteration, each thread also randomly samples $n$ negative embeddings from the cached tile to compute the gradients and update corresponding embeddings. After $N_2$ iterations, each thread randomly samples $N_1$ embeddings again to refresh the cache space. 
Since the behavior aggregation layer aggregates embeddings of a user's historical interaction items (i.e., positive items) and one user's negative items may be transformed into another user's positive items, the tiling method can also benefit the behavior aggregation layer.

\textbf{Tiling size \& refresh interval tuning.} In order to avoid manually tuning $N_1$ and $N_2$ by trial and error, we propose Algorithm~\ref{alg:tiling} to systematically determine $N_1$ and $N_2$ given an expected speedup $P$. 
Specifically,
(1) the negative sampling space of random tiling is determined by $\frac{N_1}{N_2}$ (\textcolor{crimson}{Line 2}) and affects the training results. Thus, $\frac{N_1}{N_2}$ affects the training results.
(2) We determine the latency of reading negative embeddings by estimating which level of cache can buffer a tile of embeddings (\textcolor{crimson}{Lines 5-13}).
(3) We calculate the total time of reading negative embeddings and speedup after using random tiling, and the speedup can be approximated as $\frac{N_2}{N_1}$ (\textcolor{crimson}{Lines 15-16}).
(4) We calculate the speedup of reading positive embeddings after using random tiling (\textcolor{crimson}{Line 17}).
(5) Negative and positive speedups for $\alpha, \beta$ (in percentile) of the total speedup (\textcolor{crimson}{Line 19}). In our design, we set $\alpha, \beta$ to 0.15 and 0.85, respectively.
(6) We first obtain $N_1$ via function $f_0$. The main idea of $f_0$ is to determine a suitable $N_1$ through the number of threads and the embedding size to ensure that $num\_threads \times N_1$ embeddings can be held in the L2 cache (\textcolor{crimson}{Line 21}).
(7) We can either choose the negative sampling space $I = M \times \frac{N_1}{N_2}$ or the negative speedup $neg\_speedup \approx \frac{N_2}{N_1}$ to calculate $N_2$ (\textcolor{crimson}{Lines 22-23}). 
(8) We select a smaller $N_2$ to ensure high accuracy since smaller $N_2$ larger negative sampling space (\textcolor{crimson}{Lines 24-28}).

\setlength{\textfloatsep}{0pt}
\begin{algorithm}[t]
\scriptsize\sffamily
\normalfont
\SetAlgoLined
\setcounter{AlgoLine}{0}
\SetKwComment{Comment}{\# }{}
\SetKwInOut{Input}{Inputs}
\SetKwInOut{Output}{Outputs}
\Input{$I$: \# of items, $M$: total iterations; $N_1$: tile size; $N_2$: refresh interval; $n_n$: number of negatives; $n_p$: number of positives; $r$: average positive hit ratio; $s_{l2}, s_{l3}$: L2, L3 cache size; $t_m, t_{l2}, t_{l3}$: latency of reading data from memory, L2 cache, and L3 cache; $P$: expected speedup; $\alpha, \beta$: percentage of positive, negative speedup}
\Output{$\widehat{N_1}$: optimized tile size; $\widehat{N_2}$: optimized refresh interval}
\BlankLine

\textcolor{Green}{// Negative sampling space of tiling}\par
$neg\_space \gets \frac{M}{N_2} \times N_1 = M \times \frac{N_1}{N_2}$\par

\textcolor{Green}{// Time of reading negatives using random sampling}\par
$neg\_time\_random \gets M \times n_n \times t_m$\par

\textcolor{Green}{// Estimate latency of reading cache}\par
$s_t \gets N_1 \times sizeof(embedding \; row) \times num\_threads$\par
\uIf{$s_t < s_{l2}$}{
    $t_c \gets t_{l2}$
}
\uElseIf{$s_t \geq s_{l2}\; \rm{and}\; s_t < s_{l3}$}{
    $t_c \gets t_{l3}$
}
\Else{
    $t_c \gets t_m$
}

\textcolor{Green}{// Time of reading negatives using tiling}\par
$neg\_time\_tiling \gets n_n \times \frac{M}{N_2} \times ((N_2 - N_1) \times t_c + N_1 \times t_m)$\par
$neg\_speedup \gets \frac{neg\_time\_random}{neg\_time\_tiling} = \frac{t_m}{t_c + (t_m - 1) \times \frac{N_1}{N_2}} \approx \frac{N_2}{N_1}$\par

$pos\_speedup \gets \frac{n_p \times t_m} {n_p \times r \times t_c + n_p \times (1 - r) \times t_m}$ \par

\textcolor{Green}{// Percentage of speedup}\par
$\alpha \gets \frac{pos\_speedup}{P}$
$\beta \gets \frac{neg\_speedup}{P}$

\textcolor{Green}{// Calculate $N_1$ $N_2$}\par
$N_1 \gets f_0(s_{l2},\; s_{l3},\; num\_threads,\; emb\_dim)$\par 
$N_{20} \gets \frac{M \times N_1}{I}$\par
$N_{21} \gets \frac{N_1}{\beta \times P}$\par

\eIf{$N_{20} < N_{21}$}
{
    $\widehat{N_2} \gets N_{20}$\par
}
{
    $\widehat{N_2} \gets N_{21}$\par
}
$\widehat{N_1} \gets N_1$\par


\caption{\footnotesize Proposed tuning method for tiling size \& refresh interval.}
\label{alg:tiling}
\end{algorithm}

\vspace{-1mm}
\subsection{Parallelization of Similarity Computation}
\label{subsec:similarity}
Modern CPUs are usually multi-core architectures and support the multi-threading paradigm to further exploit instruction-level parallelism. A multi-core processor typically uses a single thread in a single physical core.
In order to utilize hardware parallelism (e.g., multiple cores in CPUs, CUDA threads in GPUs) and high-performance libraries (e.g., BLAS, LAPACK), PyTorch abstracts input data into tensors (i.e., multi-dimensional matrix) and calculations into tensor operations. PyTorch-based SimpleX follows the same design philosophy. As discussed in \S\ref{subsec:efficiency}, SimpleX first concatenates embeddings, then reshapes them and adopts matrix-matrix multiplication to calculate the similarity.

However, directly adopting such a parallel computing design in CF training introduces two severe performance problems: (1) concatenating sampled embeddings into matrices and reshaping introduce expensive memory copies; and (2) normalization of embedding tensor needs writing of the generated intermediate tensor $R$ back to the memory, which causes additional memory access time. 
To conquer the above limitations and make full use of the multi-core architecture and the multi-threading paradigm in CF training, we propose a new parallel method in that different threads directly perform dot products after reading sampled user/item embeddings without concatenation and reshaping. 

\begin{figure}[t]
    \centering
     \includegraphics[width=0.95\linewidth]{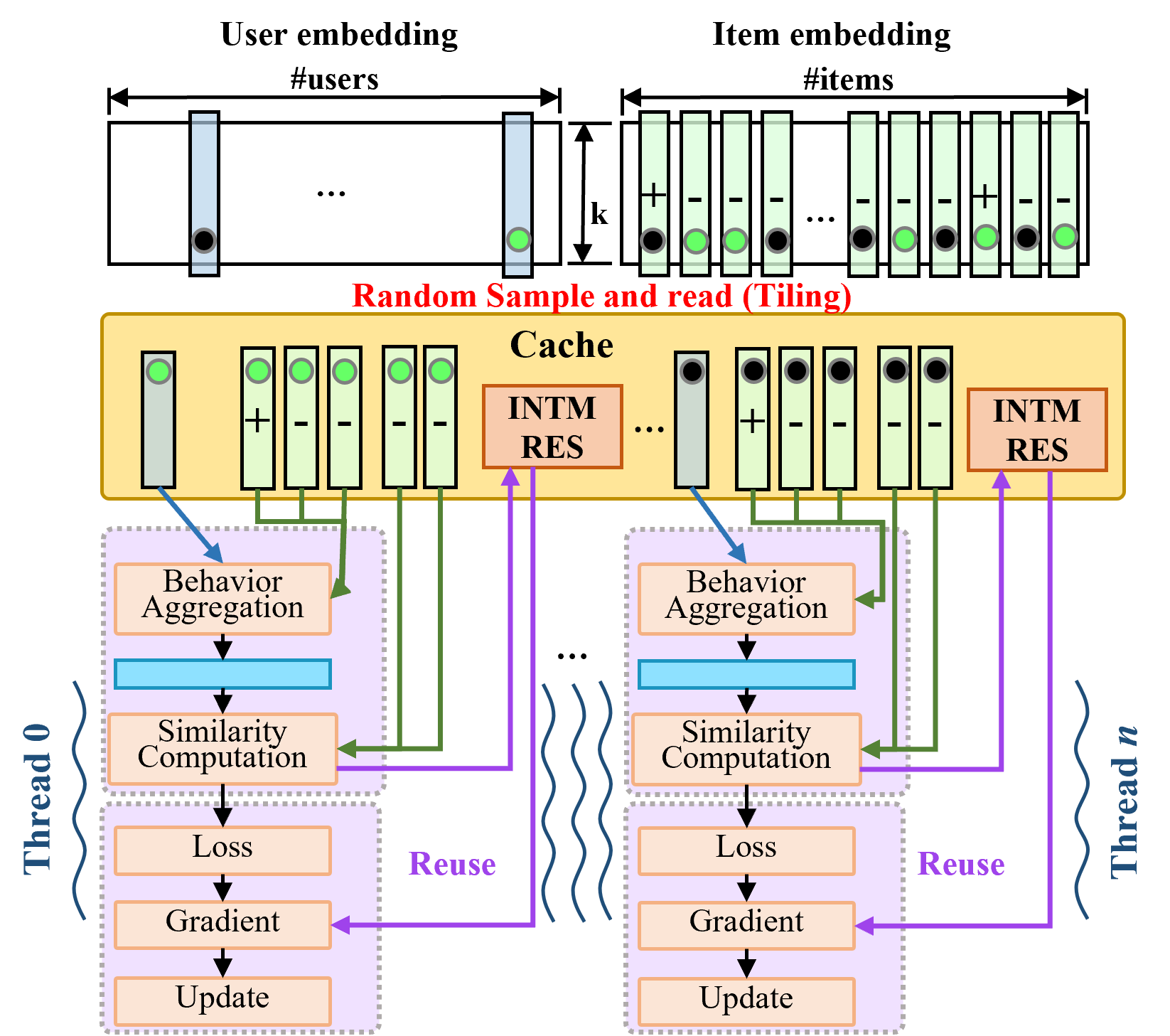}
    \caption{Overview of our training workload partition strategy. Different colored circles represent the embeddings sampled for different threads. ``+'' and ``-'' denote positive and negative embeddings, respectively.}
    \label{fig:multi_thread}
    \vspace{2mm}
\end{figure}

Figure~\ref{fig:multi_thread} depicts our proposed parallelization of similarity computation strategy. Specifically, for each iteration, each thread first fetches one user embedding $S_u$, one positive embedding $T_{pos}$, and $n$ negative embeddings $T_{neg}$. Then, each thread performs the dot product of user embedding and positive/negative embeddings $S_u \cdot T_{pos}$ or $S_u \cdot T_{neg}$. Meanwhile, to facilitate calculations of cosine similarities and reuse these embeddings, each thread also does the dot product of each embedding with itself, since $||S_u||_2 = \sqrt{S_u \cdot S_u}$, $||T_{pos}||_2 = \sqrt{T_{pos} \cdot T_{pos}}$ and $||T_{neg}||_2 = \sqrt{T_{neg} \cdot T_{neg}}$. Each thread finally generates gradients using the optimized similarity and gradient computation and then updates corresponding embeddings. 

This strategy also facilitates updating embedding matrices in a sparse fashion. Theoretically, we only need to generate the gradients of involved embeddings and update them. Note that although PyTorch allows users to set the parameter ``sparse'' to True to enable sparse gradients so as to update embeddings sparsely, it leads to worse performance as demonstrated in \S\ref{subsec:update}. By comparison, in our proposed method, different threads independently and in parallel are responsible for gradient calculations of involved embeddings. Besides, different threads can write embeddings matrices independently. Therefore, different threads are able to update these embeddings instead of updating all embeddings. 


\subsection{Aggressive Data Reuse}
\label{subsec:reuse}
With matrix factorization, the rating matrix $X$ is approximated by the matrix product of two low-rank matrices $S \in \R ^{|U| \times K}$ and $T \in \R ^{|I| \times K}$. Each row $S_u$ in $S$ can be seen as a feature vector describing a user $u$ and similarly each row $T_i$ of $T$ describes an item $i$. We need to use the feedback $X$ and a suitable loss function to train $S$ and $T$. The training procedure is (1) pick a user-item pair $(u,\; i)$ from $X$. (2) calculate the similarity $\hat{x}_{u, i}$ of the user-item pair, we can use dot product similarity or cosine similarity. We focus on cosine similarity since it delivers better training results as demonstrated in SimpleX. (3) generate loss and loss gradient using the suitable loss function. (4) do gradient backpropagation to obtain partial derivatives (gradients) of involved embeddings. (5) utilize the obtained gradients to update engaged embeddings. 
{\small
\begin{align}
\frac{\partial \hat{x}_{u, i}}{\partial S_u}
     = & \frac
     { \textstyle
        T_i \cdot \sqrt{\sum S^2_u} \sqrt{\sum T^2_i } -
        \frac{1}{2} \left(\sum S^2_u \right)^{-\frac{1}{2}} 
            \cdot 2 S_u \cdot \sqrt{\sum T^2_i} \sum S_u T_i}
    { \left( \textstyle \sqrt{\sum S^2_u} \sqrt{\sum T^2_i} \right)^{2} } \nonumber\\
   = & \frac{T_i \cdot \sum S^2_u - \sum S_u T_i \cdot S_u}{\sum S^2_u \sqrt{\sum S^2_u} \sqrt{\sum T^2_i}}
\label{equ-3}\\
\frac{\partial \hat{x}_{u, i}}{\partial T_i} =&
    \frac{S_u \cdot \sum T^2_i - \sum S_u T_i \cdot T_i}{\sum T^2_i \sqrt{\sum T^2_i} \sqrt{\sum S^2_u}} \label{equ-4}
\end{align}
}

The partial derivative of $\hat{x}_{u, i}$ with respect to the variable $S_u$ is defined in Equation (\ref{equ-3}). $\frac{\partial \hat{x}_{u, i}}{\partial S_u}$ mainly consists of $\sum S^2_u$ the sum of squares of $S_u$, $\sum T^2_i$ the sum of squares of $T_i$, and $\sum S_u T_i$ the dot product of $S_u$ and $T_i$.

We also need to calculate $\sum S^2_u$, $\sum T^2_i$, and $\sum S_u$ when calculating the cosine similarity $\hat{x}_{u, i}$ of the user-item pair in the forward phase. Thus, to avoid redundant calculation of the values of $\sum S^2_u$, $\sum T^2_i$, and $\sum S_u T_i$, we will cache the values of these variables in the forward phase to achieve data reuse.

Similarly, the partial derivative of $\hat{x}_{u, i}$ with respect to the variable $T_i$ is defined in Equation (\ref{equ-4}). $\frac{\partial \hat{x}_{u, i}}{\partial T_i}$ is also related to $\sum S^2_u$, $\sum T^2_i$, and $\sum S_u T_i$. Thus, we can reuse $\sum T^2_i$, $\sum S^2_u$, and $\sum S_u T_i$ in the calculation of $\frac{\partial \hat{x}_{u, i}}{\partial T_i}$ in the backward computation. 

\subsection{Optimized Parallel Gradient Update}
\label{subsec:aggregator}
To parallelize similarity computation (in \S\ref{subsec:similarity}), different threads independently and in parallel are responsible for similarity computation, gradient calculations, and embedding updates of involved embeddings. This parallelization strategy brings another challenge when enabling the behavior aggregation layer.

As aforementioned, the traditional MF methods only need to read one user embedding, one positive embedding, and multiple negative embeddings in each iteration, and then feed these embeddings into the model to calculate gradients and then update the corresponding embeddings. But SimpleX uses an extra behavior aggregation layer to process interacted item sequence of each user to better model user behaviors. The essence of the behavior aggregation layer is a small fully connected layer, its input/output dimension is the same as the embedding dimension.

This layer aggregates the user's embedding and embeddings of the user's historical interaction items to generate a new embedding. Then, we feed the new embedding, a positive embedding, and multiple negative embeddings, into the model to update the corresponding embeddings. The effectiveness of the behavior aggregation layer has been proven in many previous works, such as YouTubeNet \cite{covington2016deep} and ACF \cite{chen2017attentive}. It has three common aggregation choices, i.e., average pooling, self-attention, and user-attention

In \thiswork, each thread performs the training procedure independently to avoid synchronization among threads, which will degrade the overall performance. Each thread reads the weight matrix of the behavior aggregation layer to perform forward and backward computations using different training data. This training mode is similar to data parallel distributed training. We can follow asynchronous distributed stochastic gradient descent (SGD) and specify one thread as the parameter server for the global weights of the behavior aggregation layer, other threads request weights replicas from the parameter server to process a mini batch to calculate gradients and send them back to the parameter server which updates the global weights accordingly.
However, this method causes high overheads of memory and synchronization among threads due to multiple weights replicas and gradients exchange.

To solve this issue, inspired by a prior work (called Hogwild!) \cite{recht2011hogwild} that uses shared memory to hold global weights, which enables processes to access global weights without lock mechanism, we also let all threads access global weights without lock mechanism. However, Hogwild! cannot be directly applied to HEAT since Hogwild! targets to the sparse optimization problem but the optimization of the behavior aggregation layer is a dense optimization problem.
In our HEAT design, we let all threads share one weight matrix, thus each thread just holds a pointer to the weight matrix, and then generates gradients to update the weight matrix directly. The conflict will happen when one thread tries to update the weight matrix while other threads try to read/write the weight matrix since there is only one copy of the data in the memory. To alleviate the conflict, we let each thread first accumulate the gradients locally, and update the global weight matrix every $m$ iterations. 

Listing \ref{lst:alg2} describes the simplified training workflow of the behavior aggregation layer.
Specifically,
(1) we enable multi-threading processing and let aggregator\_weights be shared by all threads (\textcolor{crimson}{Line 7}).
(2) We calculate weight gradients (weights\_grad) locally and accumulate gradients to the accu\_weights\_grad (\textcolor{crimson}{Lines 13-16}).
(3) We update the global weight matrix every mini\_batch\_size (\textcolor{crimson}{Lines 17-21}). According to our experimental results, we set mini\_batch\_size to 32 to avoid accuracy drop.
\begin{lstlisting}[style=C-plus-plus, caption=Psuedocode of our behavior aggregator design., captionpos=b, label=lst:alg2, numberstyle=\tiny\color{red}, basicstyle=\scriptsize, linewidth=\textwidth]
//Input: total iteration I, init_weights0, 
//activation data act_data, outputs gradient outs_grad
//mini_batch_size
//Output: updated aggregator_weights
typedef Array<float, Dynamic, Dynamic> XMatrix
XMatrix aggregator_weights(emb_dim, init_weights0)
#pragma omp parallel shared(aggregator_weights) {
 int i_counts = 0; // iteration counts
 XMatrix weights_grad = Zero(emb_dim, emb_dim);
 XMatrix accu_weights_grad = Zero(emb_dim, emb_dim);
#pragma omp for
 for (int i=0; i<I; ++i) {
  for (int k=0; k<emb_dim; ++k) {
   weights_grad.row(k) = act_data(0, k) * outs_grad;
  }
  accu_weights_grad += weights_grad;
  if (i_counts>0 && i_counts % mini_batch_size==0) {
   weights_grad = accu_weights_grad / mini_batch_size;
   aggregator_weights -= l_r * weights_grad;
   accu_weights_grad = Zero(emb_dim, emb_dim);
  } } }
\end{lstlisting}

\section{Performance Evaluation}
\label{sec:evaluation}
In this section, we present our experimental setup and demonstrate the effectiveness of \thiswork{} compared with other solutions.

\subsection{Experimental Setup}
\label{subsec:setup}
\textbf{Datasets.} We evaluate \thiswork{} on five real-world datasets as they have been preprocessed for fairness and ease of comparison. Specifically,
(1) we perform most of our experiments on three datasets, Amazon-Books, Yelp2018, and Gowalla, which are commonly used in recent CF works \cite{chen2020revisiting, he2020lightgcn}. Amazon-Books, Yelp2018, Gowalla have 52643, 31668, 29858 users, and 91599, 38048, 40981 items, respectively.
(2) To demonstrate that \thiswork{} is affordable,
we further evaluate on two larger datasets, Goodreads Book Reviews (Goodreads) and Google Local Reviews (2018).

\textbf{Platforms.}
We perform our experiments on three types of platforms: (1) a regular memory (RM) node from the Bridges-2 supercomputer \cite{bridges-2} equipped with x86-architecture processors. Each RM node has two 64-core AMD EPYC 7742 CPUs with 32 MB L2 cache and 256 MB L3 cache;
(2) a compute node from the Ookami \cite{bari2021a64fx} cluster equipped with Fujitsu ARM A64FX processors. Each A64FX processor features 48 cores, 512-bit wide SIMD, 32 MB L2 cache, and 32 GB HBM2 memory with 1024 GB/s bandwidth; and
(3) a GPU node from the Bridges-2 supercomputer equipped with one NVIDIA Tesla 32 GB V100 GPU to perform GPU experiments. 

\textbf{Baselines.} 
SimpleX mainly consists of MF, behavior aggregation layer, and cosine contrastive loss (CCL). Note that Simplex without aggregation layer degenerates to an MF-based model. 
We compare \thiswork{} with five baselines: PyTorch-implemented MF with CCL (T-MF-CCL), TorchRec-implemented MF with CCL (R-MF-CCL), PyTorch-implemented SimpleX (T-S), TorchRec-implemented SimpleX (R-S), and CuMF\_SGD.

\textbf{Implementation details.} We implement HEAT using C++. Specifically, we implement computation kernels using Eigen \cite{eigen} for vector-dot product, and vector-matrix product. Eigen is a C++ template library for linear algebra. We use OpenMP to support our shared-memory multi-threading computation. We use Intel oneAPI C++ compiler \cite{oneapi} to compile C++ source code. We also use the Intel MKL library for BLAS operations and LAPACK operations. We use ARM C/C++ compiler \cite{arm-compiler} which provides armclang and armclang++. We link \thiswork to ARM performance library (ARMPL) to enable BLAS or LAPACK as Eigen's backend for dense matrix products. ARMPL \cite{arm-performance} provides optimized standard core math libraries such as BLAS, LAPACK, FFT, and sparse routines with OpenMP.

\begin{figure}[t]
    \centering
    \includegraphics[width=\linewidth]{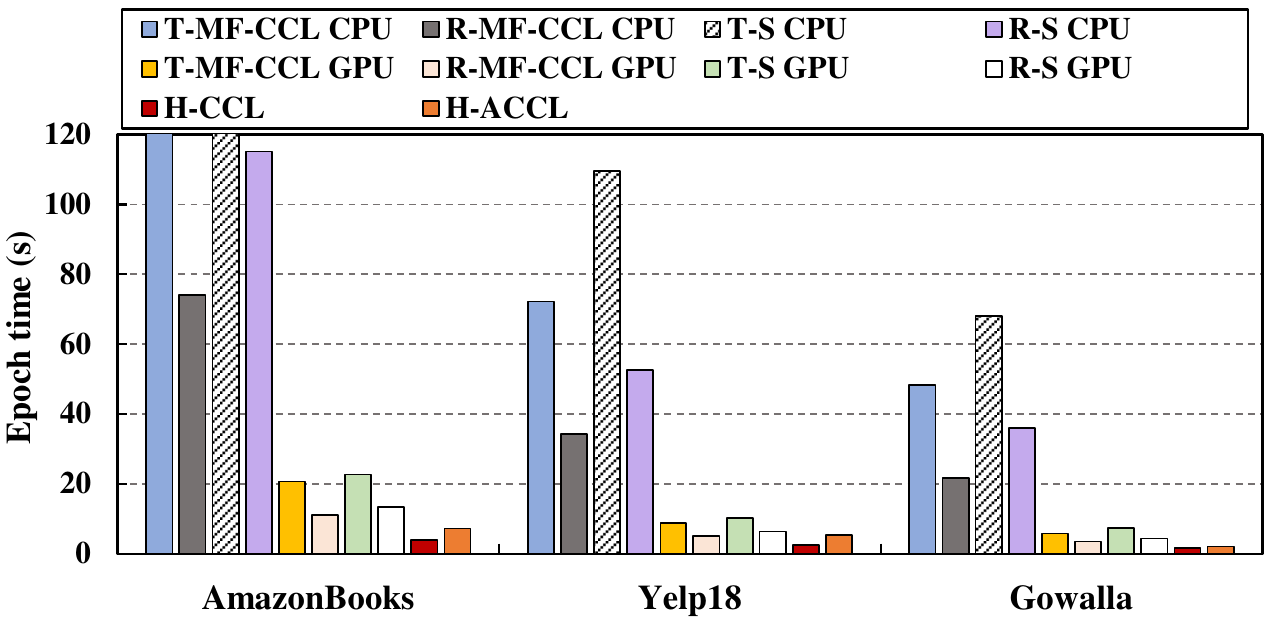}
    \caption{Comparison of epoch time between SimpleX and \thiswork{}. Simplex without aggregation layer degenerates to MF-based model.}
    \label{fig:train_speedup}
    \vspace{2mm}
\end{figure}

\subsection{Training Time}
For training epoch time, we first compare \thiswork{} with T-MF-CCL, R-MF-CCL, T-S, R-S. We run \thiswork{} on the CPU and run T-MF-CCL, R-MF-CCL, T-S, and R-S on both the CPU and GPU. For this comparison, we use the embedding dimension of 128, 64 negative samples, and 100 historical items for fairness and ease of comparison.
Moreover, we also compare \thiswork{} with CuMF\_SGD (i.e., the state-of-the-art GPU-based MF solution with high performance) and TorchRec-based MF (R-MF). For this comparison, we use the embedding dimension of 128, one negative sample, dot-product similarity, and mean square error loss because CuMF\_SGD only supports these settings.

Figure~\ref{fig:train_speedup} shows the training epoch time comparison on the CPU and GPU among T-MF-CCL, R-MF-CCL, T-S, R-S, HEAT with CCL (H-CCL), HEAT with behavior aggregation layer and CCL (H-ACCL). 
Compared with the CPU baselines, H-CCL achieves 45.2$\times$, 28.3$\times$, and 27.0$\times$ speedup over T-MF-CCL on AmazonBooks, Yelp18, and Gowalla, respectively. H-CCL also achieves 14.9$\times$ on average over R-MF-CCL. H-ACCL achieves 36.8$\times$, 20.6$\times$, and 32.1$\times$ speedup over T-S on AmazonBooks, Yelp18, and Gowalla, respectively. H-ACCL also achieves 14.3$\times$ on average over R-S.
Compared with the GPU baselines, HEAT with CCL achieves 4.5$\times$, 3.4$\times$, and 3.2$\times$ speedup over T-MF-CCL on AmazonBooks, Yelp18, and Gowalla, respectively. H-CCL also achieves 2.3$\times$ on average over R-MF-CCL. H-ACCL provides 3.2$\times$, 1.9$\times$, and 3.5$\times$ speedup over T-S on AmazonBooks, Yelp18, and Gowalla, respectively. H-ACCL achieves 1.7$\times$ speedup on average over R-S. 
We get such significant speedups because (1) we let each thread run independently and avoid synchronization between threads, (2) we aggressively reuse data in forward and backward computation to improve the performance, and (3) we only update the involved embeddings in each thread.

\begin{figure}[b]
    \centering
     \vspace{2mm}
    \includegraphics[width=\linewidth]{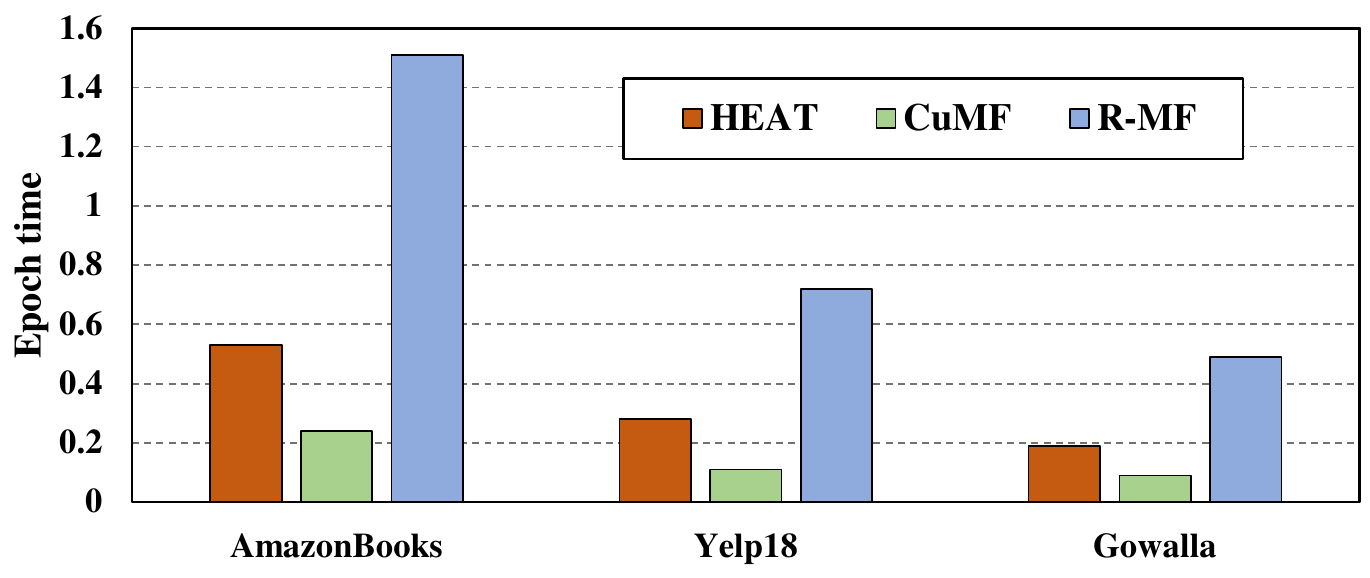}
    \caption{Comparison of epoch time among CuMF\_SGD (GPU), TorchRec (GPU), and HEAT (CPU).}
    \label{fig:cumf}
\end{figure}

\begin{figure}[t]
    \centering
    \includegraphics[width=\linewidth]{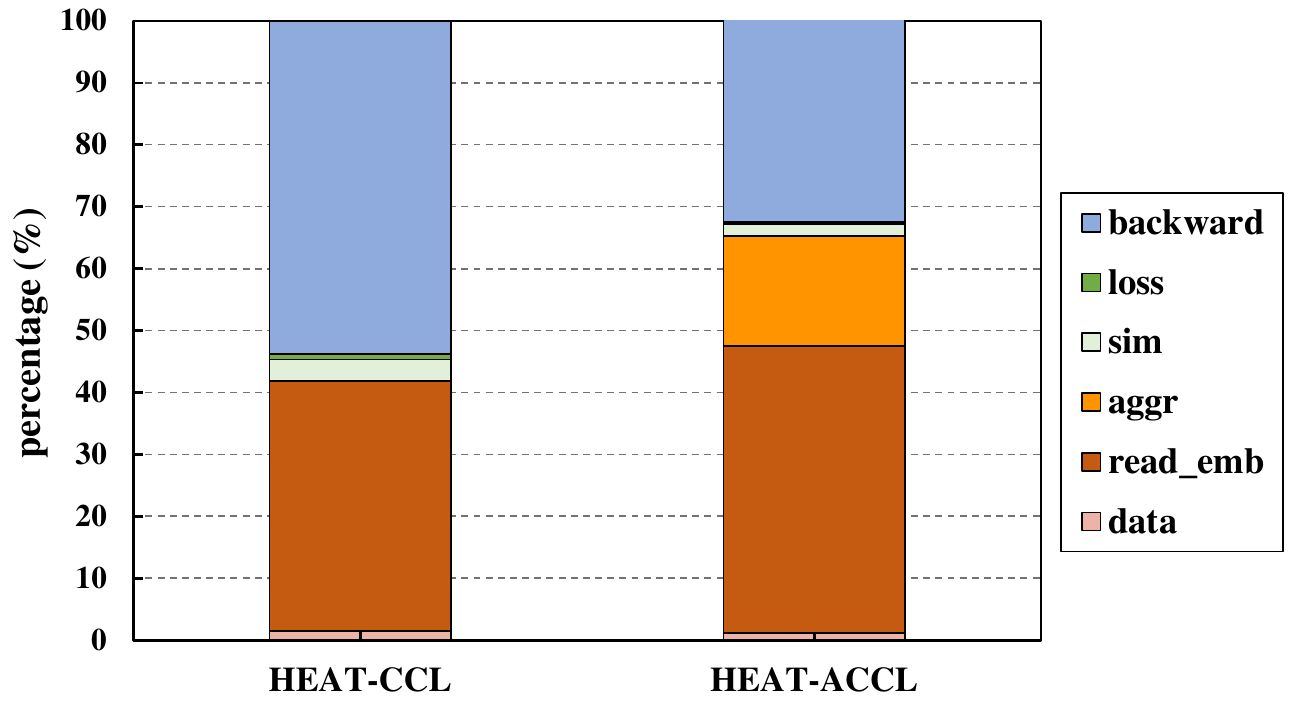}
    \caption{Performance breakdown of HEAT on CPU. Note that sim and aggr are short for similarity computation and aggregation.}
    \label{fig:breakdown}
    \vspace{2mm}
\end{figure}

Figure~\ref{fig:cumf} shows a comparison of training epoch time among CuMF\_SGD on the GPU, TorchRec-based MF on the GPU, and HEAT on the CPU. The performance of HEAT and CuMF is comparable. However, CuMF\_SGD implements the most basic CUDA-based (stochastic gradient descent) SGD solution for MF problems. CuMF\_SGD only supports basic mean squared error loss function, 1 negative sample, and sets the embedding dimension to a fixed value of 128 for performance. HEAT achieves 2.6$\times$ speedup on average over TorchRec-based MF. 

In addition, we break down the epoch time into different phases. Figure~\ref{fig:breakdown} shows that in HEAT-CCL the time of reading embeddings takes 40.4\%, which proves the necessity of our tiling strategy. Similarity computation including dot product and normalization only takes up 3.4\%, which shows our similarity computation is very efficient. Moreover, in HEAT-ACCL, the percentage of reading embeddings and aggregation reaches 46.3\% and 17.8\%, respectively, which indicates aggregation exacerbates the issue of reading embeddings and further optimization on the aggregation computation in future work.

\subsection{Training Cost}
\begin{figure}[b]
    \centering
     \vspace{4mm}
    \includegraphics[width=.95\linewidth]{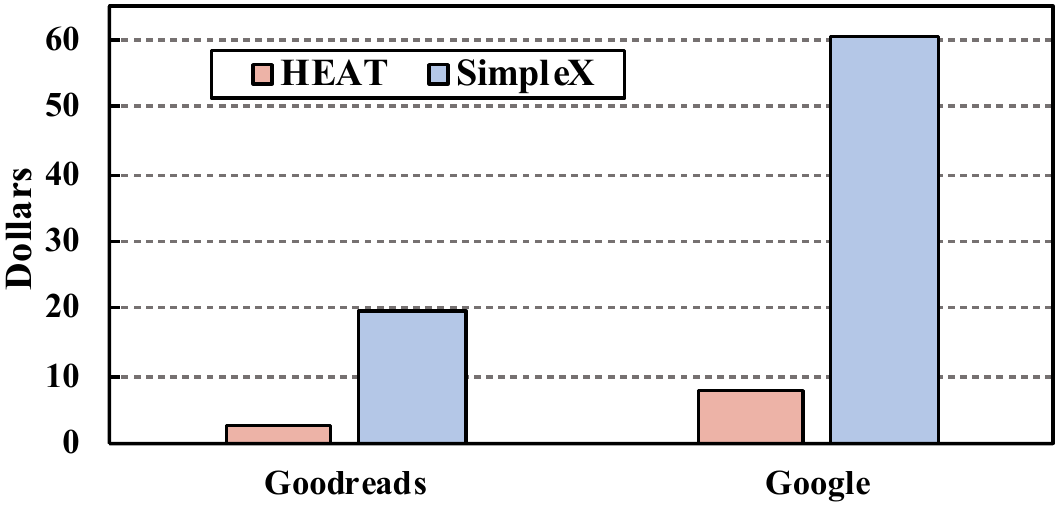}
    \caption{Comparison of total training cost (\$) for 100 epochs.}
    \label{fig:cost}
\end{figure}

Next, to demonstrate that our training system is highly affordable, we compare the training cost of our HEAT on the CPU and SimpleX on the GPU on two large datasets, i.e., Goodreads Book Reviews (Goodreads) and Google Local Reviews (2018) (Google). We use AWS p3.2xlarge instance as the GPU platform, which is equipped with one 16 GB V100 GPU. The price of p3.2xlarge is \$3.06 per hour. We need two p3.2xlarge to fit these two large datasets since each GPU has only 16 GB memory. We use AWS c5a.16xlarge as the CPU platform, which is equipped with one AMD EPYC 7R32 and 128 GB memory. The price of c5a.16xlarge is \$2.46  per hour \cite{c5a.16xlarge}. Figure~\ref{fig:cost} shows the comparison of the total training cost of the two methods for 100 epochs. Compared with SimpleX on the GPU, HEAT can reduce the cost by 7.9$\times$.

\begin{figure}[t]
    \centering
    \includegraphics[width=\linewidth]{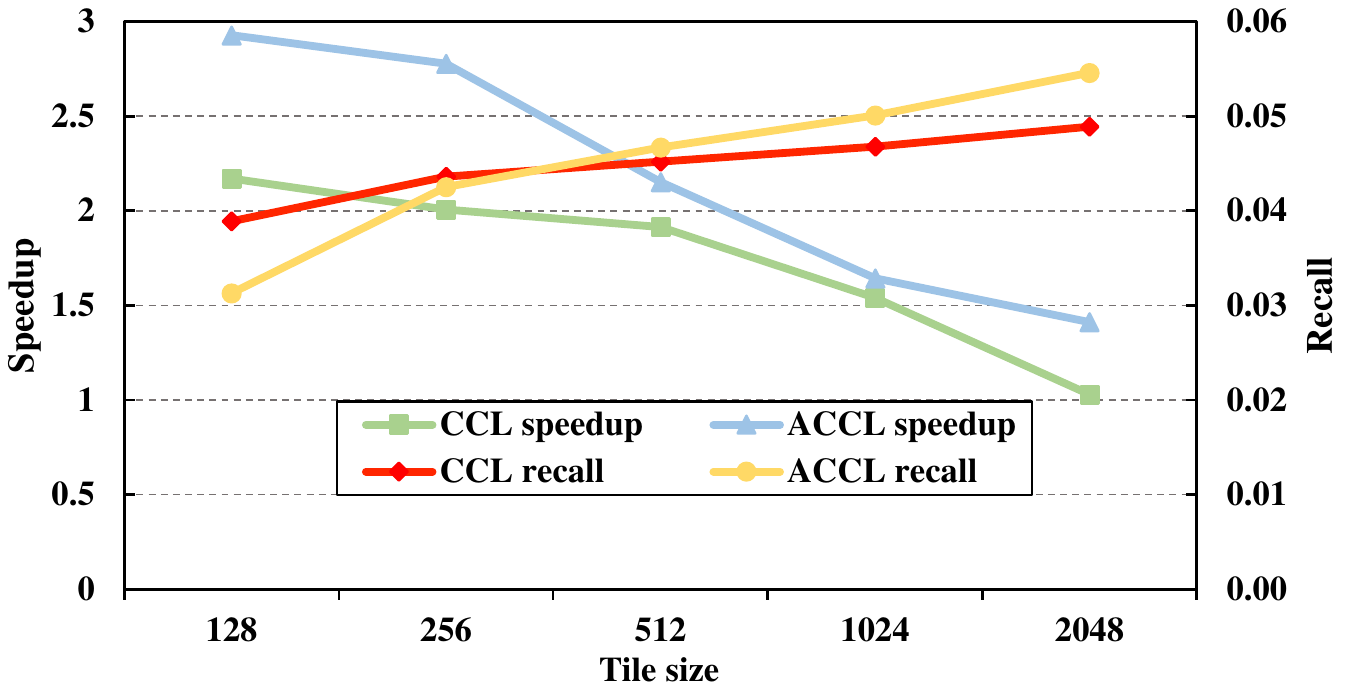}
    \caption{Speedup \& recall with different tiling sizes on AmazonBooks.}
    \label{fig:tuning_tile_size}
    \vspace{4mm}
\end{figure}

\subsection{Training Accuracy}
After that, we report the training results on different datasets using the same evaluation metrics (e.g., Recall@20 and NDCG@20) and parameter configuration as SimpleX in Table~\ref{tab:recall_random}, to demonstrate that our proposed multi-threading training system does not affect the training accuracy. Both SimpleX and HEAT's negative sampler obey the uniform distribution. 
We use the metric ``recall'', which is a widely used indicator to assess the proportion of positive samples successfully predicted by the CF model to all actually positive samples. It is calculated as $\text{Recall}= \frac{TP}{TP + FN}$, where $TP$ and $FN$ stand for true positive and false negative in the confusion matrix, respectively. ``NDCG'' is short for normalized discounted cumulative gain. The difference between the Recall@20 of HEAT and the Recall@20 of SimpleX is within $0.01$. Therefore, we can conclude that the proposed multi-threading training framework has negligible impact on training accuracy.

\begin{table*}[t]
    \caption{Comparison of training results under different frameworks and datasets.}
    \centering
    \resizebox{0.67\linewidth}{!}{
        \begin{tabular}{@{} >{\bfseries}r|rr|rr|rr}
\toprule
    \multirow{2}{*}{\sffamily Method}
    & \multicolumn{2}{c|}{\textbf{\sffamily AmazonBooks}}
    &\multicolumn{2}{c|}{\textbf{\sffamily Yelp18}}
    &\multicolumn{2}{c}{\textbf{\sffamily Gowalla}} \\
    &Recall@20 &NDCG@20 &Recall@20 &NDCG@20 &Recall@20 &NDCG@20 \\
\midrule
\midrule
MF-CCL     &0.0559  &0.0447  &0.0698  &0.0572  &0.1837  &0.1493\\
SimpleX    &0.0583  &0.0468  &0.0701  &0.0575  &0.1872  &0.1557\\
HEAT-CCL   &0.0521  &0.0416  &0.0651  &0.0548  &0.1742  &0.1413\\
HEAT-ACCL  &0.0541  &0.0429  &0.0683  &0.0561  &0.1793  &0.1457\\
\bottomrule
\end{tabular}
    }
    \label{tab:recall_random}
\end{table*}

\begin{table*}[t]
    \caption{Tiling size and refresh interval for optimal training accuracy and speedup. ``R'' and ``T'' represent random tiling sampler and random sampler, respectively.}
    \centering
    \resizebox{0.95\linewidth}{!}{
        \begin{tabular}{@{} >{\bfseries}r|rrrr|rrrr|rrrr}
\toprule
    \multirow{2}{*}{\sffamily Method}
    & \multicolumn{4}{c|}{\textbf{\sffamily AmazonBooks}}
    &\multicolumn{4}{c|}{\textbf{\sffamily Yelp18}}
    &\multicolumn{4}{c}{\textbf{\sffamily Gowalla}} \\
    &Recall@20 &Tile &Interval &Speedup &Recall@20 &Tile &Interval &Speedup &Recall@20 &Tile &Interval &Speedup \\
\midrule
\midrule
RCCL   &0.0506  &N/A   &N/A   &N/A  &0.0625  &N/A   &N/A   &N/A  &0.1691  &0.1495  &N/A  &N/A \\
RACCL  &0.0527  &N/A   &N/A   &N/A  &0.0675  &N/A   &N/A   &N/A  &0.1732  &0.1554  &N/A  &N/A \\
TCCL   &0.0498  &1024  &4096  &1.5  &0.0608  &1024  &3072  &1.8  &0.1663  &512     &4096 &1.7 \\
TACCL  &0.0518  &1024  &3072  &1.6  &0.0657  &1024  &4096  &1.5  &0.1716  &1024    &4096 &1.8 \\
\bottomrule
\end{tabular}
    }
    \label{tab:recall_tile}
\end{table*}

\begin{figure}[b]
    \centering
    \vspace{2mm}
    \includegraphics[width=\linewidth]{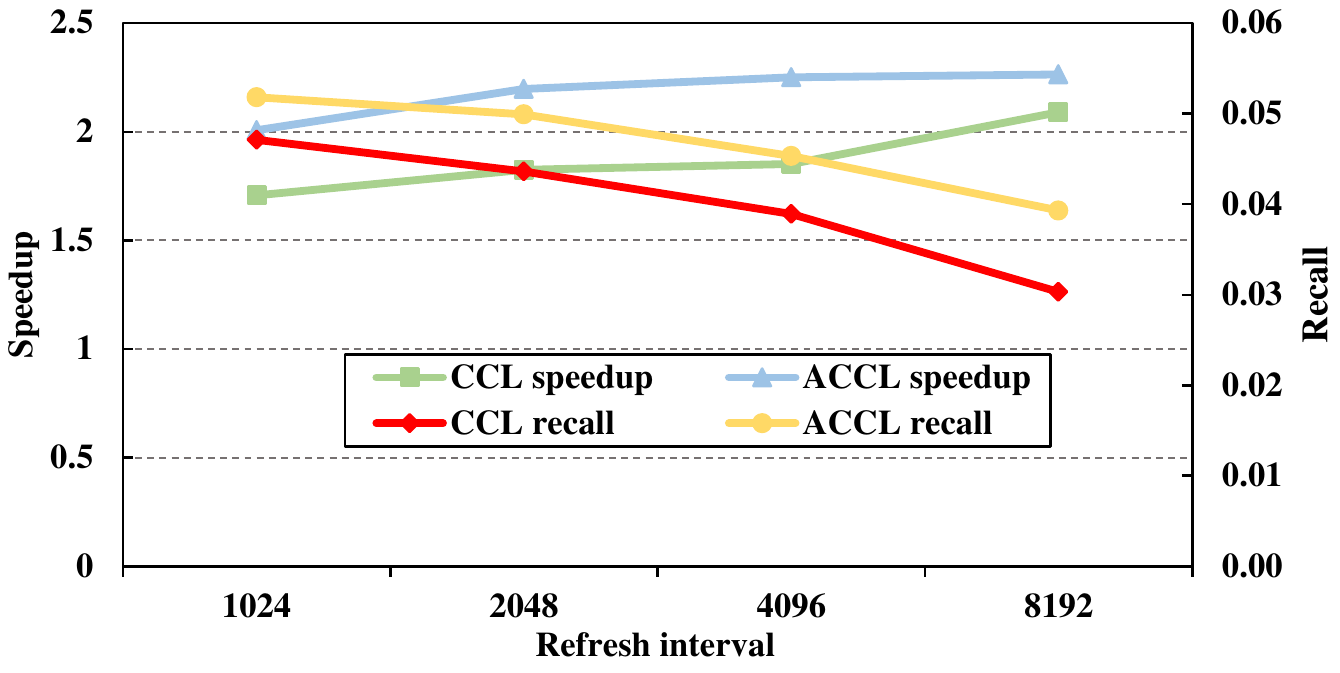}
    \caption{Speedup \& recall with different refresh intervals on AmazonBooks.}
    \label{fig:tuning_interval}
\end{figure}

\subsection{Impacts of Tiling Sizes and Refresh Intervals on Performance and Accuracy}
Furthermore, we show the effectiveness of our proposed random tiling strategy and the proposed tuning algorithm for tiling size and refresh interval. We perform experiments on AmazonBooks dataset and set the embedding dimension to 128, the number of negatives to 64, and the number of historical items to 100.

First, we show how the speedup and recall change with different tiling sizes when the refresh interval is fixed. Figure~\ref{fig:tuning_tile_size} depicts the speedup over HEAT with a random negative sampler gradually decreases with increasing tiling size. In particular, the speedup exceeds 2$\times$ when the tiling size is less than 128 because embeddings can be fully cached in the L2 cache. Meanwhile, the recall will gradually increase as the tiling size increases because the sampling space of the negative sampler increases.

Second, we show how the speedup and recall change with different refresh intervals when the tiling size is fixed. Figure~\ref{fig:tuning_interval} shows the speedup over HEAT with a random negative sampler gradually increases with increasing refresh interval. The reason is that increasing refresh interval raises the probability of data appearing in the cache, thereby reducing the time to read data. Simultaneously, the recall will gradually decrease as the refresh interval increases because the sampling space of the negative sampler decreases.
From these two experiments and the derivation of \S\ref{subsec:tiling}, we conclude that we need to adjust tiling size and refresh interval simultaneously to get the optimal accuracy and performance.

In addition, Table~\ref{tab:recall_tile} shows the tiling size and refresh interval corresponding to the optimal training results and speedup obtained by our Algorithm \ref{alg:tiling}. HEAT with the random tiling sampler delivers a 1.6$\times$ speedup on average over HEAT with the random sampler, while the recall drop is also negligible (i.e., within 0.003). 

\begin{table}[t]
    \caption{Epoch time and recall w/ and w/o local gradient accumulation.}
    \centering
    \resizebox{0.95\linewidth}{!}{
        \begin{tabular}{@{} >{\bfseries}r|rr|rr|rr}
\toprule
    \multirow{2}{*}{\sffamily Metrics}
    & \multicolumn{2}{c|}{\textbf{\sffamily AmazonBooks}}
    &\multicolumn{2}{c|}{\textbf{\sffamily Yelp18}}
    &\multicolumn{2}{c}{\textbf{\sffamily Gowalla}} \\
    &W &W/O &W &W/O &W &W/O \\
\midrule
\midrule
Epoch   &7.17    &16.92   &5.31    &9.45    &2.13    &4.96\\
Recall  &0.0527  &0.0531  &0.0675  &0.0678  &0.1732  &0.1741\\ 
\bottomrule
\end{tabular}
    }
    \vspace{2mm}
    \label{tab:aggregation}
\end{table}

\begin{figure}[b]
    \centering
    \vspace{4mm}
    \includegraphics[width=\linewidth]{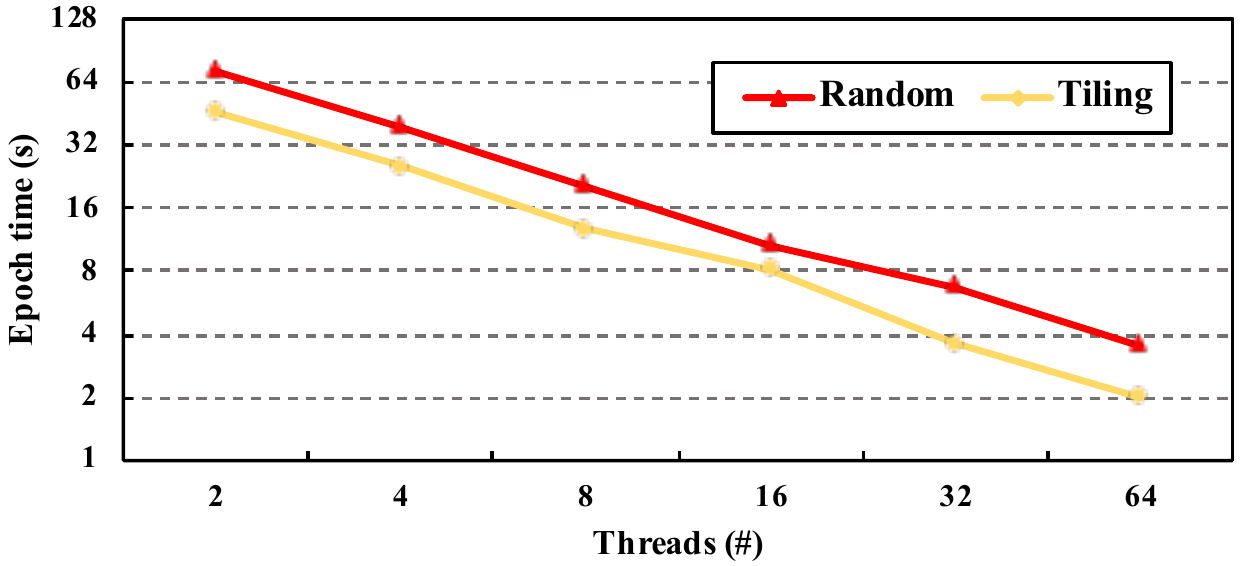}
    \caption{Scalability of \thiswork{} with original random sampler (random) and our random tiling sampler (tiling).}
    \label{fig:scalability}
\end{figure}

\subsection{Behavior Aggregation Evaluation}
To prove the proposed local gradient accumulation benefits the performance of the behavior aggregation layer, we compare the performance of HEAT with and without local gradient accumulation, as shown in Table ~\ref{tab:aggregation}. HEAT with our local gradient accumulation provides a 2.2$\times$ speedup on average due to fewer write conflicts. Moreover, its recall drop is within 0.0009. 

\subsection{Scalability Evaluation}
To demonstrate the scalability of HEAT, we choose the AmazonBooks dataset and set the embedding dimension to 128 and the number of negatives to 64. We increase the number of threads/cores from 1 to 64 (commonly used in other CF works \cite{song2020ngat4rec, wang2019neural}). Figure~\ref{fig:scalability} illustrates that the epoch time (in log-scale) decreases linearly as the number of threads increases (with the parallel efficiency of 63.7\%).
HEAT can achieve this high scalability because (1) different threads are responsible independently for the gradient calculation and embedding update in parallel, and (2) there is no need for communication and synchronization across different threads.

\subsection{Discussion of Different CPU Architectures}
To explore suitable CPU architectures for CF applications, which feature highly irregular memory access and low computation intensity, we also evaluate the performance of HEAT on an ARM-architecture processor, i.e., Fujistu A64FX, since A64FX provides 1024 GB/s bandwidth and 48 compute cores with 512-bit wide SIMD.

\begin{figure}[t]
    \centering
    \includegraphics[width=\linewidth]{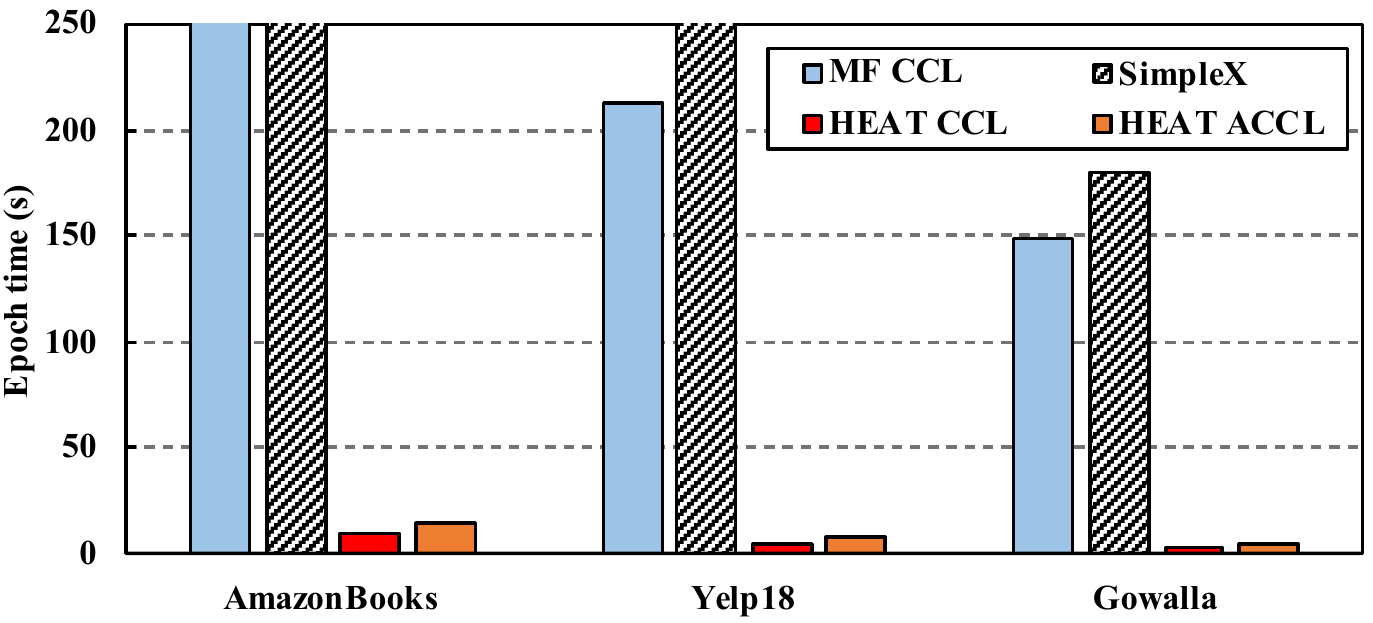}
    \caption{Comparison of training epoch time on ARM CPUs.}
    \label{fig:epoch_time_arm}
    \vspace{2mm}
\end{figure}

We first compare the overall training performance between SimpleX and HEAT on the ARM CPU. SimpleX is implemented using ARMPL-optimized PyTorch, while HEAT is compiled by armclang++ and linked to ARMPL. Figure~\ref{fig:epoch_time_arm} shows the training epoch time comparison between SimpleX and HEAT on AmazonBooks, Yelp18, and Gowalla datasets, HEAT with CCL achieves 50.4$\times$, 42.6$\times$, and 44.1$\times$ speedup over SimpleX without aggregation layer (degenerated to classic matrix factorization), respectively; HEAT with ACCL provides 41.7$\times$, 37.9$\times$, and 39.9$\times$ speedup over SimpleX with aggregation layer, respectively.

\begin{figure}[t]
    \centering
    \includegraphics[width=\linewidth]{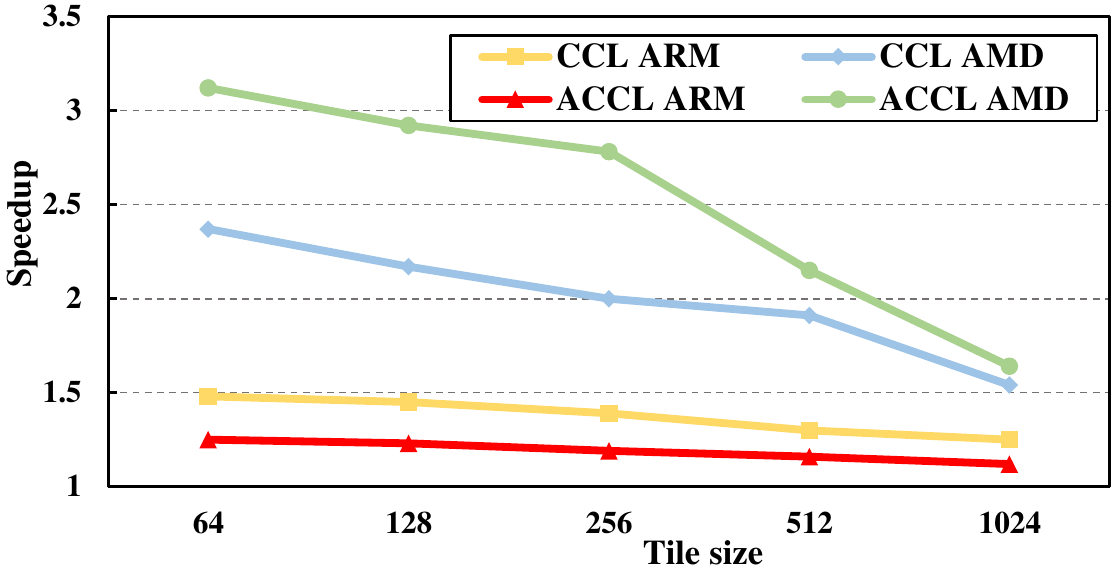}
    \caption{Comparison of tiling speedup on ARM and AMD CPUs.}
    \label{fig:arm_amd}
    \vspace{2mm}
\end{figure}

We then show a comparison of tiling speedup on the ARM CPU and on the AMD CPU in Figure~\ref{fig:arm_amd}. Specifically, our tiling optimization only provides up to 1.5 $\times$ speedup on the ARM CPU, whereas it achieves up to 3.1 $\times$ speedup on the AMD CPU. 
This is because of three reasons: (1) ARM only has two levels of caches (with a L2 cache of 32 MB), while AMD has three levels of caches (with a much larger L3 cache of 256 MB); a smaller cache leads to a higher cache miss rate. (2) Negative sampling following the uniform distribution leads to irregular memory access, which cannot give fully leverage the high memory bandwidth of HBM2. (3) The ARM CPU has fewer physical cores (i.e., 48 cores) than the AMD CPU, and the computation time takes more than 70\% of the total time, resulting in limited optimization space for tiling.

\section{Related Work}
\label{sec:related_work}
BPR \cite{rendle2012bpr} proposes a generic optimization criterion for personalized ranking via maximizing posterior estimator derived from a Bayesian analysis of the problem \cite{rendle2012bpr}. Its core idea behind is to find suitable $\Theta$ to represent parameters of an arbitrary model via maximizing posterior estimator $p(\Theta|>u) \propto p(>u|\Theta)p(\Theta)$, where $>u$ represents a user's preference. BPR concentrates on the most common scenario with implicit feedback (e.g. clicks, purchases). However, BPR uses only one negative sample, which causes inferior results for many CF models \cite{mao2021simplex}.

SimpleX \cite{mao2021simplex} investigates the impacts of the loss function, and negative sampling in CF. It demonstrates the importance of selecting an appropriate loss function and a proper number of negative samples. Inspired by contrastive loss \cite{hadsell2006dimensionality} in computer vision, SimpleX proposes a cosine contrastive loss (CCL) tailored for CF. However, SimpleX implemented its algorithm using PyTorch and did not consider the computational efficiency on either CPU or GPU.

CuMF\_SGD \cite{xie2016cumf_sgd} utilizes GPU's massive threads to update embeddings in parallel. CuMF\_SGD implemented the basic stochastic gradient descent (SGD) solution using CUDA for MF problems. However, it cannot create user-specific item ranking using the concept of positive/negative items and only supports dot-product similarity, basic mean squared error loss function and requires a fixed embedding dimension (i.e., 128) to achieve high performance. 

MSGD \cite{li2017msgd} is an MF approach for large-scale CF based recommender systems on GPUs. To parallelize SGD, MSGD removes dependencies between user and item pairs. It also splits the MF optimization objective into many separate sub-objectives. However, the optimizations of MSGD cannot be applied to multi-core CPUs because MSGD specially optimizes its parallelization approaches for coalesced memory access in GPUs. Furthermore, similar to CuMF\_SGD, it does not support sampling multiple negative items, which is crucial to the final training results. Due to the lack of source code for MSGD, we compare the performance of HEAT with SimpleX and CuMF\_SGD.

\section{Conclusion and Future Work}
\label{sec:conclusion}
In this work, we propose an efficient and affordable collaborative filtering-based recommendation training system that incorporates features of the multi-level cache and multi-threading paradigms of modern CPUs. It has a series of optimizations to address the performance issues of irregular memory accesses, unnecessary memory copies, and redundant computations. Evaluation on five widely used datasets with AMD and ARM CPUs shows that HEAT achieves up to 45.2$\times$ and 4.5$\times$ speedups over existing CPU and GPU solutions, respectively, with 7.9$\times$ cost reduction. 

In the future, we plan to first extend our work to support distributed training with rating matrix partitioning and efficient communication. 
Then, we will apply our random tiling strategy to more recommendation models such as graph neural network based CF. 

\section*{Acknowledgment}
This work was supported by the National Science Foundation (NSF) Grants OAC-2303820 and OAC-2312673. The material was also supported by the U.S. Department of Energy, Office of Science, Advanced Scientific Computing Research (ASCR), and Office of Science User Facilities, Office of Basic Energy Sciences, under Contract DEAC02-06CH11357. We gratefully acknowledge the computing resources provided by Argonne Leadership Computing Facility.
This work used the Extreme Science and Engineering Discovery Environment (XSEDE), which is supported by NSF grant number ACI-1548562. Specifically, it used the Bridges-2 system, which is supported by NSF award number ACI-1928147, at the Pittsburgh Supercomputing Center (PSC).
We would like to also thank Stony Brook Research Computing and Cyberinfrastructure, and the Institute for Advanced Computational Science at Stony Brook University for access to the innovative high-performance Ookami computing system, which was made possible by a \$5M NSF grant (\#1927880).

\newpage
\bibliographystyle{plain}
\bibliography{refs}

\end{document}